\documentclass[paper,twocolumn,twoside]{geophysics}

\usepackage{graphicx}
\usepackage{float}
\usepackage{pgfplots} 
\usepackage{amsfonts}
\usepackage{amsmath}
\usepackage{epstopdf}
\usepackage{bm}
\usepackage{caption}
\usepackage{subfig}
\usepackage{alphalph}
\usepackage{mathabx}
\usepackage{hyperref}
\usepackage{color}
\usepackage{placeins}
\usepackage{multirow}
\usepackage{enumitem}
\usepackage{array}
\usepackage{tikz}
\usepackage{multirow}

\newcolumntype{C}[1]{>{\centering}m{#1}}

\usepackage[ruled,linesnumbered]{algorithm2e}
\DeclareMathAlphabet\mathbfcal{OMS}{cmsy}{b}{n}

\date{}

\usepackage{enumitem}

\usepackage{url}

\usepackage{hyperref} 

\begin{document}

\onecolumn 

\begin{description}[labelindent=0cm,leftmargin=3cm,style=multiline]

\item[\textbf{Citation}]{Yazeed Alaudah, Motaz Alfarraj, and Ghassan AlRegib, "\textit{Structure label prediction using similarity-based retrieval and weakly supervised label mapping}", GEOPHYSICS 2019 \textbf{84:1}, V67-V79.}

\item[\textbf{DOI}]{\url{https://doi.org/10.1190/geo2018-0028.1}}

\item[\textbf{Review}]{Date of publication: 23 Aug 2018}


\item[\textbf{Bib}] {
\begin{verbatim}
@article{doi:10.1190/geo2018-0028.1,
author = {Yazeed Alaudah and Motaz Alfarraj and Ghassan AlRegib},
title = {Structure label prediction using similarity-based retrieval
and weakly supervised label mapping},
journal = {GEOPHYSICS},
volume = {84},
number = {1},
pages = {V67-V79},
year = {2019},
doi = {10.1190/geo2018-0028.1},
URL = {https://doi.org/10.1190/geo2018-0028.1},
eprint = {https://doi.org/10.1190/geo2018-0028.1},
}
\end{verbatim}
}

\item[\textbf{Contact}]{\href{mailto:alaudah@gatech.edu}{alaudah@gatech.edu}  OR \href{mailto:alregib@gatech.edu}{alregib@gatech.edu}\\ \url{http://ghassanalregib.com/} \\ }
\end{description}

\thispagestyle{empty}
\newpage
\clearpage
\setcounter{page}{1}

\twocolumn

\title{Structure Label Prediction Using Similarity-Based Retrieval and Weakly-Supervised Label Mapping}
\author{Yazeed Alaudah, Motaz Alfarraj, and Ghassan AlRegib}
\footer{Structure Label Prediction Using Similarity-Based Retrieval and Weakly-Supervised Learning}
\lefthead{Alaudah, Alfarraj, \& AlRegib}
\righthead{Structure Label Prediction}

\maketitle

\begin{abstract}
Recently, there has been significant interest in various supervised machine learning techniques that can help reduce the time and effort consumed by manual interpretation workflows. However, most successful supervised machine learning algorithms require huge amounts of annotated training data. Obtaining these labels for large seismic volumes is a very time-consuming and laborious task. We address this problem by presenting a weakly-supervised approach for predicting the labels of various seismic structures. By having an interpreter select a very small number of exemplar images for every class of subsurface structures, we use a novel similarity-based retrieval technique to extract thousands of images that contain similar subsurface structures from the seismic volume. By assuming that similar images belong to the same class, we obtain thousands of image-level labels for these images; we validate this assumption in our results section. We then introduce a novel weakly-supervised algorithm for mapping these rough image-level labels into more accurate pixel-level labels that localize the different subsurface structures within the image. This approach dramatically simplifies the process of obtaining labeled data for training supervised machine learning algorithms on seismic interpretation tasks. Using our method we generate thousands of automatically-labeled images from the Netherlands Offshore F3 block with reasonably accurate pixel-level labels. We believe this work will allow for more advances in machine learning-enabled seismic interpretation. 
\end{abstract}

\section{Introduction}

In recent years, there has been a significant interest in machine learning-based techniques for various seismic interpretation applications such as salt body delineation, fault and fracture detection, horizon extraction, and facies classification \cite[e.g.,][]{Coléou,barnes,zhen_salt,Guillen,Tao,Zhen,Silva,Jie,Ramirez, Lin2017}.

Supervised machine learning has proven to be one of the most successful machine learning paradigms. By definition, supervised machine learning techniques require labels to perform training. However obtaining labels for large volumes of seismic data is a very demanding task, especially when the size of the data is in the tens or hundreds of gigabytes. Furthermore, while the amount of data is growing continuously, the ability of human experts to manually label large quantities of data is insufficient. Therefore, the time and effort required to manually annotate large amounts of training data can often exceed the time and effort \textit{saved} by techniques based on automated machine learning workflows. This is where, we believe, \emph{similarity-based retrieval} and \emph{weakly-supervised learning} can play a significant role in overcoming this problem. 

\begin{figure}
    \centering
    \includegraphics[width=0.5\textwidth]{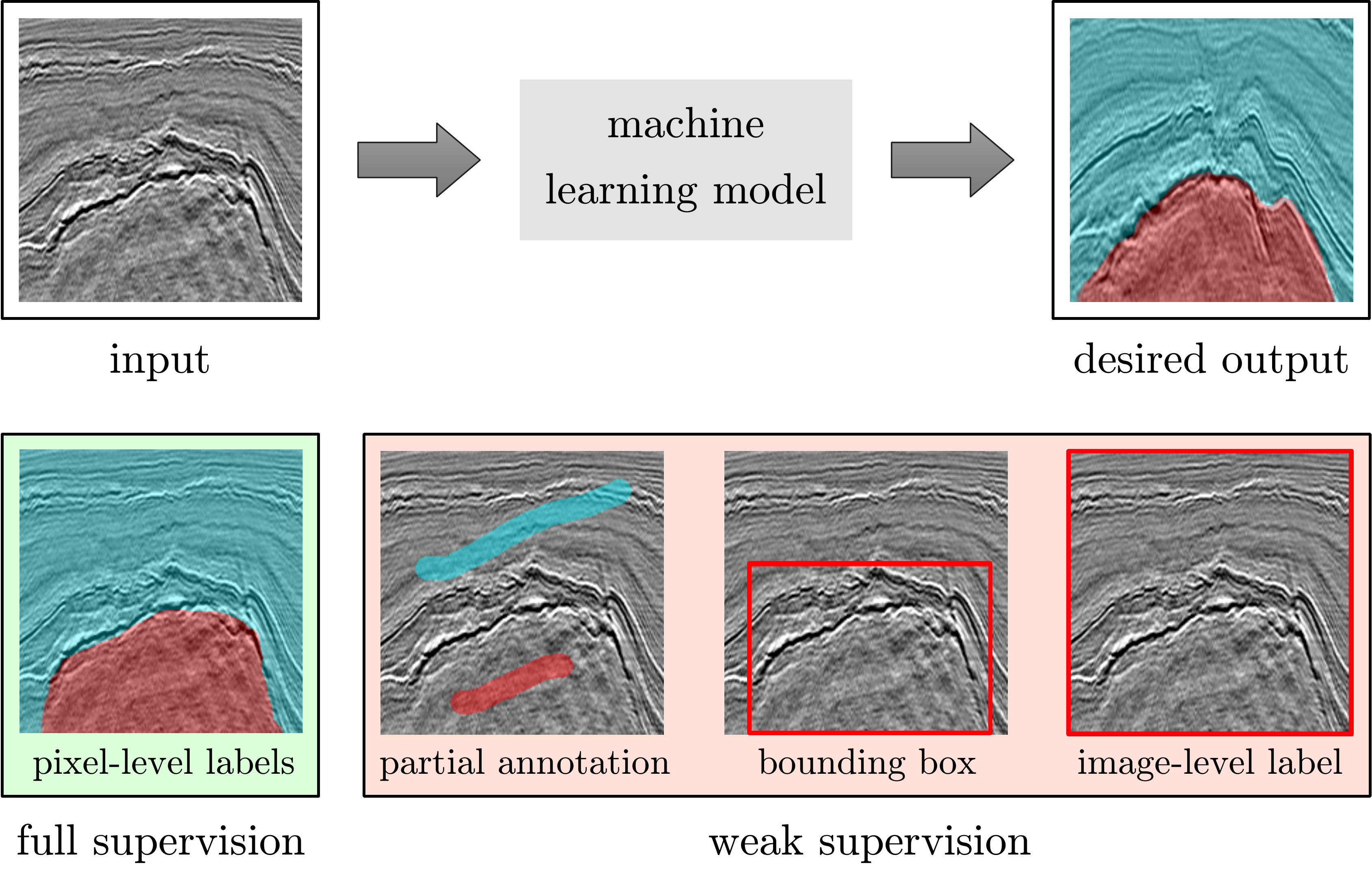}
    \caption{The difference between full supervision and weak supervision. At the bottom row from left to right: pixel-level labels, partial labels, bounding box, and an image level label. Red denotes salt body, and cyan indicates everything else.}
    \label{fig:weakly-supervised}
\end{figure}

Similarity-based retrieval is a widely used technique in content-based image retrieval (CBIR) applications \cite[]{zhou2017recent}. It utilizes metrics designed to measure the similarity of two images, to retrieve visually similar images from a large database of images. On the other hand, weakly-supervised learning is a machine learning paradigm where the model is trained using examples that are only partially annotated or labeled
\cite[]{Torresani2014}. Figure \ref{fig:weakly-supervised} explains this further. Assume we would like to train a simple machine learning model to classify salt bodies. The model takes an input image similar to the one in the figure and produces the desired output shown, where red denotes salt body, and cyan denotes everything else. To train a \textit{fully}-supervised model, we would need training labels that are \textit{fully} annotated. In this case, this means we need pixel-level labels for all pixels in the training images. If the interpreter only partially labeled the training images, or provided a bounding box, or just provided an image-level label indicating whether the training image contained a salt body or not, then our trained machine learning model would be a \textit{weakly}-supervised one, and our labels are considered \textit{weak} labels.   

In this work, we show that it is possible to \textit{predict} thousands of high-quality pixel-level labels for training supervised machine learning models for seismic interpretation with very minimal input from the interpreter. As few as one or two exemplar images are required for each subsurface structure of interest such as faults or salt domes. This is achieved using an unsupervised similarity-based seismic image retrieval technique to extract thousands of images with visually similar subsurface structures. A weakly-supervised matrix-factorization based technique is then used to learn the common structures and features between these images, and then map their image-level labels into pixel-level labels that can be effectively used to train powerful fully supervised machine learning models. 

\raggedbottom

Preliminary elements of this work were published in \cite[]{alfarraj_MMSP,yazeed_WSL_EAGE17,yazeed_WSL_SEG17}. To our knowledge, other than this work, no other work attempts to address this issue in the field of seismic interpretation. Nevertheless, we overview the related literature in similarity-based retrieval and weakly-supervised labeling in the following subsections.

\subsubsection{Similarity-Based Retrieval}

Image similarity measures are functions that quantify the similarity between a pair of images. The definition of image similarity can vary depending on the application. Generic similarity or distance measures, such as peak signal-to-noise ratio (PSNR) and Euclidean distance, assume a pixel-to-pixel correspondence between images. These measures treat every pixel independently. Structural similarity (SSIM) is a similarity measure that improves upon the pixel-to-pixel metrics by capturing local image structure using low-level local statistics in the spatial domain (S-SSIM)\cite[]{ssim} or the complex wavelet domain (CW-SSIM)\cite[]{cwssim}.
These measures are often used for applications such as image denoising where the pixel-to-pixel correspondence is justified. Another class of similarity measures is content-based similarity measures which quantify the similarity between the contents of the images without making any assumptions about the correspondence of the locations on the content. Such measures are often used for applications like content-based image retrieval in which the goal is to find images that contain similar visual content to a query image. A particular class of content-based similarity measures is texture-based similarity measures which compare the textured content of images using texture analysis. A well-known example of such measures is the structural texture similarity measure (STSIM) \cite[]{stsim} which uses subband statistics and correlations in a multiscale frequency decomposition, namely, the steerable pyramid. Another example of such multiscale decomposition is the curvelet transform \cite[]{c2curve}. It provides an efficient way of representing images with high directional content. \cite{c2curve} have shown that images that contain geometrically regular edges are more compactly represented by a curvelet rather than a wavelet decomposition. This is especially true for seismic data, where the wavefronts lie mainly along smooth curves. A few texture-based similarity measures based on curvelet coefficients have been proposed in the literature for texture and seismic image similarity. \cite{hasan} proposed a seismic image similarity measure based on adaptive curvelets. \cite{long2015seisim} proposed combining the texture similarity metric in \cite[]{stsim} with seismic discontinuity maps to evaluate the similarity of seismic images. \cite{icip2015} then proposed a method based on histograms of curvelet coefficients, while \cite{alfarraj_MMSP} proposed a method based on the singular values of curvelet coefficients, truncated using the effective rank \cite[]{roy2007effective}.

\subsubsection{Weakly-Supervised Labeling}

The vast majority of the visual data nowadays is unlabeled or weakly labeled. Lately, there has been considerable interest in weakly-supervised methods for labeling natural images. Several weakly-supervised methods proposed recently are based on matrix factorization techniques. For example, \cite{sonmf} proposed a framework for clustering images retrieved from a reference dataset using sparse and orthogonal non-negative matrix factorization. However, this method was proposed for image clustering and does not apply to pixel-level labeling. Other researchers have considered techniques based on non-negative matrix factorization to infer the pixel-level labels of pre-segmented regions (known as superpixels) within different images \cite[]{niu2015weakly,lu2017learning}. Furthermore, \cite{Cofactorization} proposed a non-negative matrix co-factorization based approach that jointly learns a discriminative dictionary and a linear classifier that classifies features from segmented images into different classes. These methods are all applied on \textit{features} extracted from superpixels and not the actual pixels in the image. This considerably limits the resolution of the resulting labels. Furthermore, most of these approaches assume or require that the image is segmented into semantic regions that represent different classes. This is not the case in our work.  
 
More recently, driven by the success of convolutional neural networks (CNNs) in various computer vision tasks, many researchers have proposed weakly-supervised labeling methods based on CNNs \cite[]{papandreou2015weakly,pinheiro2015image,hong2015decoupled,zhang2015weakly,pathak2015constrained,hou2016mining,kim2016deconvolutional,wei2017stc,kim2017two}. These methods, however, require a significant amount of training data to be effective, and usually do not localize various objects accurately without requiring other post-processing steps. Our proposed approach, based on non-negative matrix factorization, does not require any training or post-processing. In fact, very few exemplar images per class are required for our approach to give satisfactory results. As few as one or two exemplars per class can be sufficient to obtain a large number of image-level labeled data using similarity-based retrieval. These image-level labels that we obtain are then directly mapped into pixel-level labels without using any segmentation or any feature extraction technique.  

In summary, our main contributions in this work are as follows:
\begin{itemize}
\item We introduce a similarity-based image retrieval framework to extract large numbers of images from a 3D seismic volume with similar subsurface structures to exemplar images chosen by an interpreter.
\item We propose a weakly-supervised algorithm, based on non-negative matrix factorization, to learn a mapping from image-level labels into pixel-level labels. 
\item We apply this approach to images extracted from the Netherlands Offshore F3 block and generate thousands of pixel-level labeled images for three classes of subsurface structures: chaotic layers, faults, and salt domes.  
\end{itemize}

\noindent The rest of the paper is organized as follows: first, we introduce the proposed approach including the similarity-based retrieval and the weakly-supervised label mapping. Then, we show various results that illustrate the effectiveness of our similarity-based retrieval method, as well as the weakly-supervised label mapping. We then discuss these results and how our method can be improved, and finally, conclude the paper. 

\section{Proposed Method}

\label{theory}

\subsection{Overall Workflow}
\begin{figure*}[t]
\centering
\includegraphics[width = 0.99\textwidth]{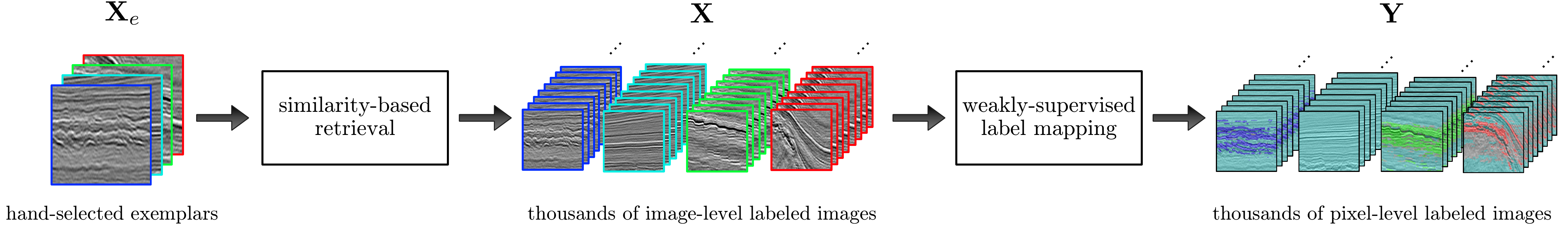}
\caption{General workflow of the proposed method. Colored image boundaries indicate the image-level labels. Blue, cyan, green, and red denote the \texttt{chaotic}, \texttt{other}, \texttt{faults}, and \texttt{salt dome} classes respectively.}
\label{fig:overall_block_diagram}
\end{figure*}

\begin{figure}[h]
  \centering
    \subfloat[\texttt{chaotic}]{\includegraphics[width=0.16\textwidth]{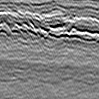}} \hspace{0.0\textwidth}
    \subfloat[\texttt{faults}]{\includegraphics[width=0.16\textwidth]{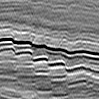}}
    \hspace{0.0\textwidth}
    \subfloat[\texttt{other 1}]{\includegraphics[width=0.16\textwidth]{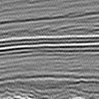}}
    \hspace{0.0\textwidth}
    \subfloat[\texttt{other 2}]{\includegraphics[width=0.16\textwidth]{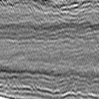}}
    \hspace{0.0\textwidth}
    \subfloat[\texttt{salt dome 1}]{\includegraphics[width=0.16\textwidth]{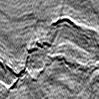}} \hspace{0.0\textwidth}
    \subfloat[\texttt{salt dome 2}]{\includegraphics[width=0.16\textwidth]{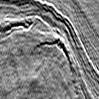}} 
  \caption{The exemplar images of each class of subsurface structures that were used to retrieve the images in this work. One exemplar image was used for \texttt{chaotic} and \texttt{fault}, and two exemplars were used for \texttt{other} and \texttt{salt dome}. These images are of size $99 \times 99$ pixels and were obtained from the Netherlands Offshore F3 Block \cite[]{F3_data}.}
  \label{fig:exemplars}
\end{figure}

The overall workflow of the proposed method is shown in Figure \ref{fig:overall_block_diagram}. First, the interpreter hand-selects a few images $\mathbf{X}_e = [\mathbf{x}_1, \mathbf{x}_2, \cdots,
\mathbf{x}_{N_e}]$ such as those in Figure \ref{fig:exemplars}, to exemplify each class of subsurface structures. These can be as few as one image per class if the visual features of that class are uncomplicated. A similarity-based retrieval method is then used to extract a very large number of images that contain similar subsurface structures. We assume that most-similar images to the exemplar belong to the same class as that of the exemplar image. We validate this assumption later in the results section. At this stage, all these images have image-level labels. A weakly-supervised mapping, based on non-negative matrix factorization, is then used to map these image-level labels into pixels to obtain the final results. A high-level workflow of the proposed approach is depicted in Figure \ref{fig:overall_block_diagram}.  The remaining parts of this section explain the similarity-based retrieval and the weakly-supervised label mapping modules in detail. 

\subsection{Similarity-Based Retrieval}

Image similarity measures are used to evaluate the similarity of the visual content between two images. They take two images as input and return a value, often in the range $[0,1]$, that indicates the level of similarity between the two images. A higher value indicates higher similarity. These measures are often used to search for images within large visual datasets. Here, we present the similarity measures we use to capture the similarity of seismic images \cite[]{alfarraj_MMSP} and then describe how this metric is used to retrieve images that contain similar subsurface structures. This similarity measure works by computing the similarity between two vectors containing the singular values of the curvelet decomposition of the two images, trimmed adaptively using effective rank approximation. We will explain this in detail in the following subsections.

\subsubsection{The Curvelet Transform}



\begin{figure}[t]
\centering
\subfloat[Frequency viewpoint]{\includegraphics[width=0.5\linewidth]{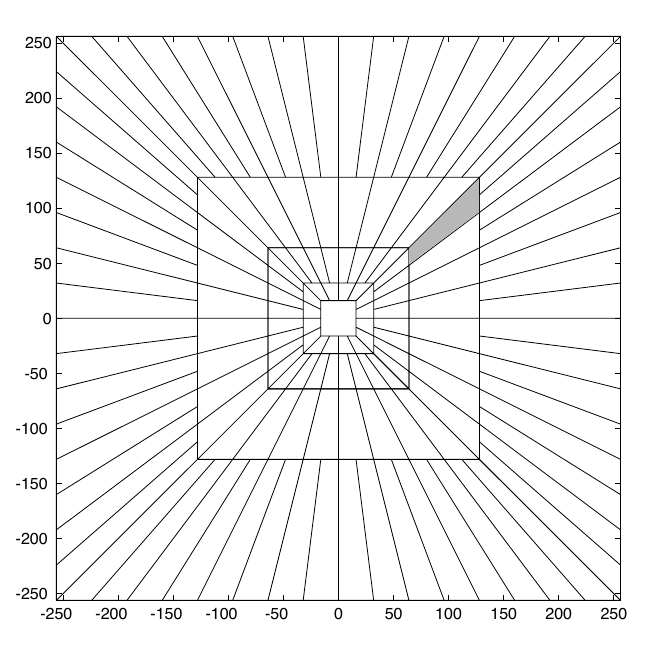}}

\subfloat[Spatial viewpoint]{\includegraphics[width = 0.42\linewidth]{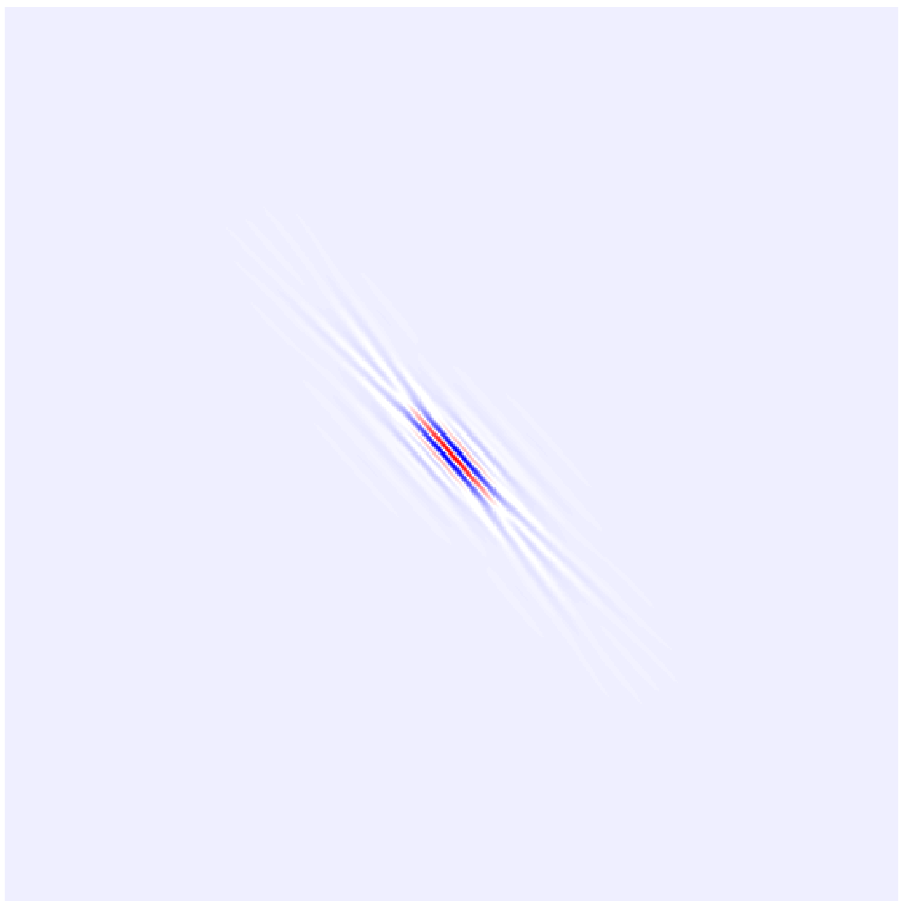}}
\caption{Frequency and spatial viewpoints of a curvelet wedge. Adapted from \cite{candes05}. Copyright \copyright 2006 Society for Industrial and Applied Mathematics. Reprinted with permission. All rights reserved.}
\label{fig:curvelet}
\end{figure}

The curvelet transform is a directional multiscale decomposition first introduced by Candes \textit{et al.}\cite[]{candes04,candes05}. It provides an efficient way of representing images with high directional content. \cite{c2curve} have shown that the curvelet transform provides an optimally sparse representation for curve-like structures, such as seismic reflectors, when compared to wavelets. 

The curvelet transform works by taking the 2D fast Fourier transform of an image (2D FFT) and then dividing the plane into multiple scales and orientations as is shown in Figure \ref{fig:curvelet}(a). The total number of scales in the curvelet tiling, $J$, depends on the size on the image, and is given by
\begin{equation}\label{nScales}
J =  \lceil \log_2 \min (N_1, N_2) -3 \rceil ,
\end{equation} 
where $N_1$ and $N_2$ are the number of pixels in vertical and horizontal directions, respectively; and $\lceil \cdot \rceil$ is the ceiling function. The number of orientations at scale $j\geqslant 1$, $K(j)$, is given by:
\begin{equation}
K(j) = 16 \times 2^{\lceil (j-1)/2 \rceil}.
\end{equation}
\noindent For scale $j=0$, there is only one orientation. Curvelet coefficients are then generated by taking the inverse FFT for each wedge (such as the one highlighted in Figure \ref{fig:curvelet}) after multiplying it by a smooth band pass filter. Since the FFT of real images is symmetric around the origin, only two consecutive quadrants of the Fourier spectrum are necessary for obtaining the curvelet coefficients.  Figure \ref{fig:curvelet} shows the spatial and frequency representations of a curvelet wedge.

\subsubsection{Feature Extraction and Similarity Measurement}
\begin{figure*}[]
\centering
\includegraphics[width=0.75\textwidth]{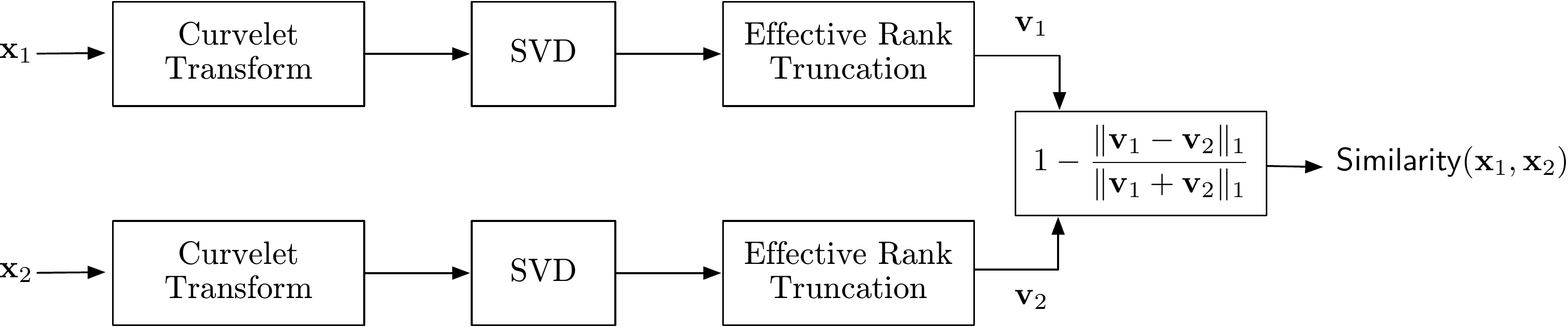}
\caption{Workflow of the similarity measure proposed in \cite{alfarraj_MMSP}.}
\label{fig:retrieval_block_diagram}
\end{figure*}

To compute the similarity between two images, we first compute feature vectors for each image and then we arrive at the similarity value by comparing the corresponding feature vectors of two images. The feature vector of a given image is a collection of all effective singular values for all scales and orientations of the curvelet coefficients of an image. The overall workflow for computing the similarity is depicted in Figure \ref{fig:retrieval_block_diagram}.  

To obtain a feature vector for a grayscale image, $\mathbf{x}_i$, we first apply the curvelet transform, and compute its coefficients for all scales, $j=\{1,2,\dots,J\}$, and orientations, $k =\{1,2, \dots, K(j)\}$. Then, the singular values of these curvelet coefficients are calculated as $\bm{\sigma}_{[j,k]} = [\sigma_1,\dots,\sigma_L]^T$ where $\sigma_1 \geq \sigma_2 \hdots \geq \sigma_L$ and $L$ is the smallest dimension of the coefficients matrix. 

Ideally, if the rank of a matrix is $r$, only the first $r$ singular values are non-zero. However, when we consider the singular value decomposition (SVD) on images that are subject to different types of noise, the number of non-zero singular values is greater than $r$. In most cases, none of the singular values are exactly zero; even for a rank-deficient matrix. \cite{roy2007effective} proposed the \emph{effective rank} as a method to estimate the actual rank of a matrix by estimating its effective dimensionality. To calculate the effective rank, we first compute the normalized singular values as

\begin{equation}\label{SVdistribution}
p_i = \frac{\sigma_i}{\|\bm{\sigma}_{[j,k]}\|_1} ~\text{for } i = 1,\dots, L,
\end{equation}
where $\|\cdot \|_1$ is the $\ell_1$ norm. Then, the effective rank is calculated as a function of the entropy of the singular value distribution defined in equation \ref{SVdistribution}, that is
\begin{equation}\label{eq:effectiveRank}
\mathsf{EffectiveRank} = \exp{\left(-\sum_{i=1}^{L} p_i \log p_i\right)},
\end{equation} 
resulting in a real number less than or equal to $L$ with equality if and only if all singular values are equal. 

For each set of curvelet coefficients, the $\mathsf{EffectiveRank}$, is calculated as in equation \ref{eq:effectiveRank}. A new vector of \textit{effective} singular values is formed by keeping the first $\lfloor \mathsf{EffectiveRank} \rfloor$ singular values, where $\lfloor \cdot  \rfloor$ denotes the floor function. The remaining singular values are set to 0. In other words, for scale $j$ and orientation $k$, we form the vector $\hat{\bm{\sigma}}_{[j,k]} = [\sigma_1,\dots,\sigma_{\lfloor \mathsf{EffectiveRank} \rfloor},0,\dots,0]$. The overall feature vector of image $\mathbf{x}_i$ is then obtained by concatenating all $\hat{\bm{\sigma}}_{[j,k]}$ for all scales and half the number of orientations,
\begin{equation}\label{eq:featVec}
\mathbf{v}_i = [\hat{\bm{\sigma}}_{[1,1]},\hat{\bm{\sigma}}_{[2,1]}, \dots, \hat{\bm{\sigma}}_{[2,K(2)/2]},\hat{\bm{\sigma}}_{[3,1]}\dots,\hat{\bm{\sigma}}_{[J,1]}].
\end{equation}

\noindent The similarity between two images, $\mathbf{x}_1$ and $\mathbf{x}_2$, is then computed as 
\begin{equation}
\mathsf{Similarity} (\mathbf{x}_1,\mathbf{x}_2) = 1-\frac{\|\mathbf{v}_1-\mathbf{v}_2\|_1}{\|\mathbf{v}_1+\mathbf{v}_2\|_1}.
\end{equation}

\noindent Where, $\mathbf{v}_1$ and $\mathbf{v}_2$ are the feature vectors corresponding to $\mathbf{x}_1$ and $\mathbf{x}_2$. Since the singular values are non-negative by definition, the resulting similarity value is in the range $[0,1]$ with a value closer to $1$ indicating higher similarity.

\subsubsection{Similarity-Based Retrieval}

By ranking the retrieved images based on their similarity to a reference image, we can extract thousands of images that resemble exemplar images chosen by an interpreter. In our work, we extract thousands of images extracted from random locations within the Netherlands Offshore F3 block \cite[]{F3_data}. We then use similarity-based retrieval to retrieve the top $M$ images that are most similar to the exemplar images shown in \ref{fig:exemplars}. The value of $M$ is selected by the interpreter such that the $M$ retrieved images can be reasonably assumed to belong to the same class as the reference exemplar image. Later in the results section, we show that our approach is robust to wrongly-retrieved images.

\subsection{Weakly-Supervised Label Mapping}

Once we have obtained image-level labels, we use a weakly-supervised learning approach, based on non-negative matrix factorization, to map these image-level labels into pixel-level labels. This mapping is a weakly supervised one since every image $\mathbf{x}_i  \in \mathbb{R}_+^{n\times m}$ has only one label as opposed to $ n\times m$ labels (i.e., one label per pixel). In the remainder of this section, we will describe this weakly-supervised label mapping in detail. 

Given the image-level labeled images, $\{ \mathbf{x}_1, \mathbf{x}_2, \cdots, \mathbf{x}_{N_s} \} $, we vectorize them to construct the data matrix $\mathbf{X} \in \mathbb{R}_+^{N_p \times N_s}$ such that each image is a column in $\mathbf{X}$, where $N_p = n \times m$ is the number of pixels in each image, and $N_s$ is the total number of images.  

\subsubsection{Non-Negative Matrix Factorization}
\label{NMF_overview}

Non-negative Matrix Factorization (NMF) \cite[]{paatero1994positive,Lee2000} is a commonly used matrix factorization technique that is closely related to many unsupervised machine learning techniques such as $k$-means and spectral clustering \cite[]{Turkmen2015a,Ding2005}. NMF  decomposes a non-negative matrix $\mathbf{X} \in \mathbb{R}_+^{N_p \times N_s}$ into the product of two lower-rank matrices $\mathbf{W} \in \mathbb{R}_+^{N_p \times N_f}$, and $\mathbf{H} \in \mathbb{R}_+^{N_f \times N_s}$ such that both $\mathbf{W}$ and $\mathbf{H}$ are non-negative, and $N_f < \min(N_p, N_s)$. In other words we have,
\begin{equation}
\mathbf{X} \approx \mathbf{W}\mathbf{H}.  
\end{equation}

\noindent In our work, the matrix $\mathbf{X}$ represents a data matrix where each column is a single image in vector form. The data matrix  $\mathbf{X}$ has $N_s$ such images, each of which is a vector of length $N_p$. Here, we use $N_p$, $N_s$, and $N_f$ to denote the number of pixels, the number of samples, and the number of features (or the \textit{rank} of $\mathbf{X}$) respectively. NMF factorizes the data matrix $\mathbf{X}$ into two non-negative matrices, a basis matrix $\mathbf{W}$ and a coefficient matrix $\mathbf{H}$. In clustering terms, the columns of $\mathbf{W}$ represent $N_f$ number of clusters in the data, whereas the columns of $\mathbf{H}$ represent the memberships of each of the images to the different clusters in the data. Here, the clusters can represent different seismic structures like salt domes, faults, or horizons.   
    
The regular NMF problem does not have a closed-form solution, and is typically solved by minimizing the following objective function 
\begin{equation}\label{NMFopt}
\underset{\mathbf{W},\mathbf{H}}{\arg\min}  || \mathbf{X} - \mathbf{W}\mathbf{H} ||_F^2  ~~~\textrm{s.t.}  \mathbf{W} \geq 0 ~\mathrm{and}~ \mathbf{H} \geq 0,
\end{equation}
\noindent where, $||\cdot||_F$ is the Frobenius norm, and $\geq$ is an element-wise inequality. \cite{Lee2001} proposed an efficient method of solving this problem using multiplicative update rules, and proved that they converge to a local minima. 

\subsubsection{Sparsity and Orthogonality Constraints}

 \cite{Lee2000} showed that NMF can be used to learn a parts-based representation of the data, where each feature would represent a localized ``part" of the data. In practice, this is rarely achieved using the formulation in equation \ref{NMFopt}. To remedy this, we initialize the feature matrix $\mathbf{W}^0$ using $k$-means separately on the different classes in the data matrix $\mathbf{X}$. This ensures that each feature $\mathbf{w}_i$ in the matrix $\mathbf{W}$ corresponds to a single class. We then impose a sparsity constraint on these initial features such that the sparsity of every feature $\mathbf{w}_i$ in matrix $\mathbf{W}^0$ satisfies  

\begin{equation} \label{eq:sparsity}
\rho(\mathbf{w}_i) = \frac{\sqrt{N_p} - { ||\mathbf{w_i}||_1 \over ||\mathbf{w_i}||_2}}{\sqrt{N_p} - 1},
\end{equation}

\noindent where $\rho(\cdot)$ indicates the sparsity of a vector. This value is always between zero and one, with higher values indicating higher sparsity. To enforce this sparsity constraint, we follow the algorithm proposed by \cite{Hoyer2004}. Additionally, we would expect that our features in $\mathbf{W}$ only represent a few images. In other words, it is unlikely that the same feature will be present in a large number of the images across different classes. To enforce this expectation, we impose an orthogonality constraint on the coefficients matrix $\mathbf{H}$. We also add two regularization terms on $\mathbf{W}$ and $\mathbf{H}$. The problem then becomes

\begin{equation}
\begin{aligned}\label{SONMF}
\underset{\mathbf{W},\mathbf{H}}{\arg\min}& ~ || \mathbf{X} - \mathbf{W}\mathbf{H} ||_F^2  +  \gamma||\mathbf{H}\mathbf{H}^T - \mathbf{I}||_F^2 + \lambda_1 ||\mathbf{W}||_F^2 \\ + &\lambda_2||\mathbf{H}||_F^2  
~~\textrm{s.t.}~  \mathbf{W}\geq 0, \mathbf{H} \geq 0 ~\mathrm{and}~  \rho(\mathbf{w}_i) = \rho_w,
\end{aligned}
\end{equation}

\noindent where matrix $\mathbf{I} \in \mathbb{R}_+^{N_f \times N_f}$ is an identity matrix. The values $\gamma$, $\lambda_1$, and $\lambda_2$ are regularization constants, and $\rho_w$ is the desired sparsity level.

\subsubsection{Multiplicative Update Rules}

Instead of solving the problem in equation \ref{SONMF} for both $\mathbf{W}$ and $\mathbf{H}$, we decouple this problem into two separate sub-problems. The first, 
\begin{equation}\label{eq:WObjective}
\underset{\mathbf{W}}{\arg\min} ~ || \mathbf{X} - \mathbf{W}\mathbf{H} ||_F^2  +  \lambda_1 ||\mathbf{W}||_F^2 ~~ \textrm{s.t.}  \mathbf{W} \geq 0 , \rho(\mathbf{w}_i) = \rho_w,
\end{equation}
is solved for $\mathbf{W}$ while $\mathbf{H}$ is held constant. Then the second, 
\begin{equation}\label{eq:HObjective}
\underset{\mathbf{H}}{\arg\min} ~ || \mathbf{X} - \mathbf{W}\mathbf{H} ||_F^2  +  \gamma||\mathbf{H}\mathbf{H}^T - \mathbf{I}||_F^2 + \lambda_2||\mathbf{H}||_F^2  
~~\textrm{s.t.}  \mathbf{H} \geq 0,
\end{equation}
is solved for $\mathbf{H}$ while $\mathbf{W}$ is held constant. We use gradient descent to derive the following multiplicative update rules for $\mathbf{W}$:

\begin{equation}\label{WMUR}
\mathbf{W}^{t+1} = \frac{ \mathbf{W}^{t} \odot({\mathbf{X}}{\mathbf{H}^t}^T)_{ij}}{(\mathbf{W}^t\mathbf{H}^t{\mathbf{H}^{t}}^T +\lambda_1\mathbf{W}^{t})_{ij}},
\end{equation}

\noindent and for $\mathbf{H}$:
\begin{equation}\label{HMUR}
\mathbf{H}^{t+1} =  \frac{\mathbf{H}^t \odot \big({\mathbf{W}^{t+1}}^T {\mathbf{X}} + \gamma \mathbf{H}^{t}\big)_{ij}}{ {(\mathbf{W}^{t+1}}^T\mathbf{W}^{t+1}\mathbf{H}^{t} +\lambda_2\mathbf{H}^{t} + \gamma\mathbf{H}^t{\mathbf{H}^{t}}^T\mathbf{H}^t  )_{ij}}.
\end{equation}

\noindent
Here, $\odot$ represents element-wise multiplication, and the superscript indicates the iteration number. These multiplicative update rules (MURs) are applied successively until both $\mathbf{W}$ and $\mathbf{H}$ converge. As we show in Appendix B, these multiplicative update rules are a special case of gradient descent with an automatic step size selection. One advantage of using these MURs is the guaranteed non-negativity of $\mathbf{W}$ and $\mathbf{H}$ when they are initialized with non-negative values. A detailed derivation of these MURs is shown in Appendix B. 

\subsubsection{Extracting the Labels}

Once $\mathbf{W}$ and $\mathbf{H}$ have converged, each column of $\mathbf{H}$, $\mathbf{h}_n$, indicates the features used to construct the  $n^\text{th}$ image. Since every feature in $\mathbf{W}$ should correspond to a single class, we can predict the label of each pixel in the image by knowing which features are used to represent it. In other words, we can map the coefficients in  $\mathbf{h}_n$ to the seismic structures that make up the image. Thus for image $\mathbf{x}_n$ we can obtain      
\begin{equation}\label{7}
\mathbf{L}_n  = \mathbf{W}(\mathbf{Q}\odot (\mathbf{h}_n \mathbf{1}^T)) ~~~~~ \forall n = [1,\cdots,N_s],
\end{equation}
\noindent where $\mathbf{1}$ is a column vector of ones of length $N_l$, and $\mathbf{Q} \in \{0,1\}^{N_f \times N_l}$ is a cluster membership matrix such that the element $\mathbf{Q}(i,j) = 1$ if the feature $\mathbf{w}_i$ belongs to structure $j$. The matrix $\mathbf{Q}$ is used to encode our knowledge of the image-level labels, and how the matrix $\mathbf{W}$ was initialized. The resulting matrix, $\mathbf{L}_n\in \mathbb{R}_+^{N_p \times N_l}$ shows the likelihood of each seismic structure for each pixel in the image. Then, the pixel-level labels for image $\mathbf{x}_n$ correspond to the seismic structure given by 

\begin{equation}\label{8}
\mathbf{y}_n(i) = \arg \underset{j}{\max}  ~ \mathbf{L}_{n}(i,j) ~~~~ \forall i = [1, \cdots, N_p],
\end{equation}

\noindent where $\mathbf{L}_{n}(i,j)$ denotes the element in the $i^\text{th}$ row and $j^\text{th}$ column of matrix $\mathbf{L}_{n}$. However, due to the nature of the weakly-supervised mapping of the labels, there is an element of uncertainty in the mapping. Since the features $\mathbf{w}_i$ are sparse, some pixels in an image $\mathbf{x}_n$ may not have a feature that accurately represents all the pixels within it. These pixels typically end up being represented as a weighted sum of a large number of different features, often from different classes and having small coefficients. This leads to noisy labeling results. To remedy this, we introduce a new \texttt{uncertain} class that contains pixels with uncertain labels. We define our confidence, $\mathbf{c}_n \in \mathbb{R}^{N_p}$, in the predicted label of every pixel in the image $\mathbf{x}_n$ as
\begin{equation} \label{eqn:confidence}
\mathbf{c}_n(i) = \max_j ~ \mathbf{L}_n(i,j) ~~~~ \forall i = [1, \cdots, N_p].
\end{equation}
We can then assign any pixel whose confidence is less than a threshold $\tau$ to the \texttt{uncertain} class, denoted as class $0$

\begin{equation}
    {{\mathbf{y}}_n}_{(\mathbf{c} < \tau)} = 0.
\end{equation}

\noindent Once we obtain the pixel-level labels $\mathbf{y}$ for each image, we apply a $3 \times 3$ median filter to clear any noisy labels and get the final labeling result for that image. We do this for all $N_s$ images and concatenate the results to construct the pixel-level labels matrix $\mathbf{Y} \in \mathbb{Z}^{N_p \times N_s} $ that contains the final pixel-level labels for all the images in the data matrix
\begin{equation}
    \mathbf{Y} = [{\mathbf{y}}_1, {\mathbf{y}}_2, \cdots , {\mathbf{y}}_{N_s}].
\end{equation}

\FloatBarrier
\section{Results}

We present our results in two subsections corresponding to the two main modules of our workflow shown in Figure \ref{fig:overall_block_diagram}. First, we show results that demonstrate the effectiveness of our similarity measure and validate our assumption that all retrieved images should have the same image-level label as the query image. Then we show sample results of the weakly-supervised label mapping and demonstrate its effectiveness in accurately localizing the various seismic structures we have used in this study.

\label{results}

\subsection{Similarity-Based Retrieval}
\label{results:retreival}

To evaluate the performance of our similarity measure compared to other measures, we devise two experiments. Namely, retrieval and clustering. We do both experiments directly on the similarity matrices of the various measures. These matrices contain the similarity values between all pairs of images in a dataset for a specific measure. For example, for a dataset that contains $N_s$ images divided into $N_l$ classes, the size of the similarity matrix $\mathbf{S}$ is $N_s\times N_s$, where $\mathbf{S}(i,j)$ is the similarity between $\mathbf{x}_i$ and $\mathbf{x}_j$, i.e. $\mathbf{S}(i,j) = \mathsf{Similarity} (\mathbf{x}_i,\mathbf{x}_j)$. The $i^\text{th}$ row of $ \mathbf{S}$ represents the similarity values of all images in the dataset compared to $\mathbf{x}_i$. Note that in these experiments we convert distance metrics, such as Euclidean distance and the method proposed by \cite{icip2015}, to similarity measures by normalizing each row of $\mathbf{S}$ by the maximum value of that row and subtracting it from 1. Hence, the diagonal entries of $\mathbf{S}$ are all 1. 

Throughout the similarity-based retrieval experiments, we use the LANDMASS-2 dataset\footnote{\href{http://cegp.ece.gatech.edu/codedata/landmass/}{https://ghassanalregib.com/landmass/}} \cite[]{LANDMASS} which is comprised of 4000 images of size $99 \times 99$ pixels, with their values normalized to be between 0 and 1. These images were extracted from the Netherlands Offshore F3 block and divided equally into four classes according to their dominant structure. The classes are \texttt{horizon}, \texttt{chaotic}, \texttt{fault} and \texttt{salt dome}. In the following subsections, we will discuss the two experiments and their results in detail. 

In our experiments, we compare the performance of the method we described \cite[]{alfarraj_MMSP} to different similarity and distance measures. The following measures were used in the experiments:
\begin{enumerate}[nolistsep]
\item Euclidean distance 
\item CW-SSIM with default parameters \cite[]{cwssim}
\item STSIM-1 and STSIM-2 with 4 scales and 8 orientations \cite[]{stsim}
\item SeiSIM with 4 scales and 8 orientations \cite[]{long2015seisim} 
\item Curvelet-based distance measure \cite[]{icip2015}
\end{enumerate}

Figure \ref{fig:roc} shows the receiver operating characteristic (ROC) curves of the similarity measures listed above. The retrieval results are summarized in Table \ref{table:retrieval}. The metrics used to evaluate the performance of different methods are detailed in Appendix A.  The results show that the similarity measure proposed in \cite[]{alfarraj_MMSP} is the best performing in all the different metrics we used. 

To further analyze the retrieval results, we show in Figure \ref{fig:precision} the Precision @$M$ curves for each of the four classes in the dataset using the best performing similarity measure, in addition to Precision @$M$ for the entire dataset. The cyan curve represents the \texttt{horizon} class, and it shows that the precision is 1 for all values of $M$. This is mainly due to the simplicity of structures that appear in \texttt{horizon} images. On the other hand, the curves of the other classes drop at different rates depending on the complexity of their structures. For example, images of \texttt{fault} class have a different number of faults of different scales and dipping angles, which makes them the most complicated class in the dataset. Hence, the \texttt{fault} precision curve drops at a faster rate than those of other classes. We show the curve for the combined images of all classes in black.

\input{other/ROC_curve}
\input{other/precision_curve}

\begin{table*}[ht]
\centering
\caption{Performance of different similarity metrics for similarity-based retrieval and clustering}
\label{table:retrieval}
\begin{tabular}{l || ccc| c ||}
& \multicolumn{3}{c|}{Retrieval} & Clustering\\\cline{2-5}
Metric & RA & MAP & AUC & Rand Index \\
\hline
Euclidean distance      & 0.345 & 0.394 & 0.515 & 0.394\\
CW-SSIM\cite[]{ssim}    & 0.721 & 0.806 & 0.858 & 0.870\\
STSIM-1\cite[]{stsim}   & 0.867 & 0.926 & 0.966 & 0.895\\
STSIM-2\cite[]{stsim}   & 0.855 & 0.910 & 0.964 & 0.877\\
SeiSIM\cite[]{long2015seisim}   & 0.819 & 0.886 & 0.945 & 0.888\\
\cite{icip2015}         & 0.896 & 0.949 & 0.978 & 0.905\\
\cite{alfarraj_MMSP}    & \textbf{0.911 }& \textbf{0.954} & \textbf{0.983} & \textbf{0.970}\\
\hline

\end{tabular}
%
\end{table*}

To further assess the performance of the similarity measures listed above on seismic data, we set up a clustering experiment using the similarity matrix obtained previously. First, the images in the dataset are projected into a 2-dimensional Euclidean subspace based on their similarity, such that the distance between images in the projection subspace is inversely proportional to their similarity values. The projection is done using classical multidimensional scaling (MDS) \cite[]{mds}. Then, the projected data points are clustered into four clusters using the $k$-means algorithm.

Since clustering is an unsupervised method, the clustering algorithm clusters the images according to their measured similarity. Therefore, the clusters do not necessarily correspond to classes; unless the similarity measure has a good discriminative power. Therefore, one can use the clustering results to quantify the goodness of the similarity measure. To evaluate the clustering performance, we compute the rand index (explained in Appendix A) which is a measure of the similarity between two data clusterings. We report the rand index results for different similarity measures in Table \ref{table:retrieval}. The results of the clustering experiment further validate our conclusion from the retrieval experiment that our method is superior to other methods in the literature. 

Also, we show the two-dimensional projection of the data using our method in Figure \ref{fig:mdscale}. The figure shows that using the similarity values to project the dataset into a lower dimensional subspace produces clusters that are almost linearly separable. \texttt{Horizon} and \texttt{salt dome} classes are separated well from all other classes. However, the \texttt{fault} and \texttt{chaotic} classes slightly overlap. It is important to mention that this is only a 2D projection of the data and that the data is more easily separated in a higher-dimensional space. These results suggest that the similarity measure can be used to discriminate the different classes of seismic images with high accuracy.

\input{other/mds_figure}

\subsection{Weakly-Supervised Label Mapping}

We apply our similarity measure on the Netherlands Offshore F3 block \cite[]{F3_data} to retrieve $M= 500$ images for each of the four classes, \texttt{chaotic}, \texttt{other},  \texttt{fault}, and \texttt{salt dome}.  We put these images in vector form and use them to construct matrix $\mathbf{X}$. We then apply the $k$-means clustering algorithm on each class separately and use the results to initialize matrix $\mathbf{W}^0$ after we impose the sparsity constraint in equation \ref{eq:sparsity}. The coefficients matrix $\mathbf{H}^0$ is initialized with uniform random numbers in the range $[0,1]$. We empirically select the values of $\lambda_1, \lambda_2$ and $\gamma$ as $0.1$, $0.5$, and $5$ respectively. The sparsity of the initial features $\rho_w$ is set to $0.4$. Additionally, the confidence threshold $\tau$ is set to $0.001$.
 
\begin{figure}[h]
\resizebox{\columnwidth}{!}{
\centering
\definecolor{mycolor1}{rgb}{0.00000,0.44700,0.74100}%
\definecolor{mycolor2}{rgb}{0.85000,0.32500,0.09800}%
\definecolor{mycolor3}{rgb}{0.92900,0.69400,0.12500}%
\definecolor{mycolor4}{rgb}{0.49400,0.18400,0.55600}%
\definecolor{mycolor5}{rgb}{0.46600,0.67400,0.18800}%
\begin{tikzpicture}

\begin{axis}[%
at={(0.772in,0.481in)},
scale only axis,
xmin=0,
xmax=0.2,
xlabel style={font=\color{white!15!black}},
xlabel={Percentage of images in $\mathbf{X}$ that are replaced with wrongly-labeled images},
xtick={0,0.5,0.10,0.15,0.2},
ymin=0,
ymax=1,
ylabel style={font=\color{white!15!black}},
ylabel={Relative Performance},
axis background/.style={fill=white},
xmajorgrids,
ymajorgrids,
legend style={at={(0.98,0.02)},anchor=south east,legend cell align=left, align=left, draw=white!15!black}
]

\addplot [color=mycolor1, line width=1.5pt,mark=+, mark options={solid, mycolor1}]
  table[row sep=crcr]{%
0	1\\
0.01	0.939815651525453\\
0.02	0.931644870632943\\
0.03	0.923876123922627\\
0.04	0.929604108594182\\
0.05	0.911047615568433\\
0.0600000000000001	0.907002669801429\\
0.0700000000000001	0.916543740281317\\
0.08	0.90205277452755\\
0.09	0.897929633967155\\
0.1	0.882801263810024\\
0.11	0.895545755781115\\
0.12	0.881574181696474\\
0.13	0.865359449875065\\
0.14	0.872738896175585\\
0.15	0.843366020434258\\
0.16	0.861656867778139\\
0.17	0.83518987801719\\
0.18	0.832804461504641\\
0.19	0.833277272322219\\
0.2	0.807856119120281\\
0.21	0.806620603480177\\
0.22	0.799769456890621\\
0.23	0.792141785685308\\
0.24	0.764565666770367\\
0.25	0.778971403563858\\
};
\addlegendentry{$k=$150}

\addplot [color=mycolor2, line width=1.5pt,mark=o, mark options={solid, mycolor2}]
  table[row sep=crcr]{%
0	1\\
0.01	0.990027747541281\\
0.02	0.983486522311843\\
0.03	0.979152040588441\\
0.04	0.977048997942063\\
0.05	0.970922286324707\\
0.0600000000000001	0.968098427994523\\
0.0700000000000001	0.971555820677018\\
0.08	0.955922146803453\\
0.09	0.948161050185366\\
0.1	0.941918846057756\\
0.11	0.940378356508027\\
0.12	0.91618844718111\\
0.13	0.935824820749293\\
0.14	0.924728558315602\\
0.15	0.901943956261564\\
0.16	0.901026702838242\\
0.17	0.879148137329292\\
0.18	0.888834377421093\\
0.19	0.874544985952872\\
0.2	0.873374297425773\\
0.21	0.857139626637154\\
0.22	0.843996041951785\\
0.23	0.846738248181163\\
0.24	0.836245572497036\\
0.25	0.800912864823967\\
};
\addlegendentry{$k=$200}

\addplot [color=mycolor3, line width=1.5pt,mark=asterisk, mark options={solid, mycolor3}]
  table[row sep=crcr]{%
0	1\\
0.01	0.996793717509797\\
0.02	0.99412836520203\\
0.03	0.990793004483916\\
0.04	0.985402777722752\\
0.05	0.983594153884406\\
0.0600000000000001	0.980025173657062\\
0.0700000000000001	0.982774963749657\\
0.08	0.975865135885851\\
0.09	0.973028792822084\\
0.1	0.968007646923279\\
0.11	0.961792238027547\\
0.12	0.953260254479032\\
0.13	0.948445994555075\\
0.14	0.934584584933407\\
0.15	0.931318077659541\\
0.16	0.925211090798094\\
0.17	0.922666922880253\\
0.18	0.914399477750349\\
0.19	0.902733459840667\\
0.2	0.901922142066332\\
0.21	0.878255476036427\\
0.22	0.8842948888577\\
0.23	0.848245477590795\\
0.24	0.852939114550054\\
0.25	0.858725137688418\\
};
\addlegendentry{$k=$250}

\addplot [color=mycolor4, line width=1.5pt,mark=x, mark options={solid, mycolor4}]
  table[row sep=crcr]{%
0	1\\
0.01	0.998109023680306\\
0.02	0.996247645861472\\
0.03	0.993216373326997\\
0.04	0.990259450422814\\
0.05	0.989928092986604\\
0.0600000000000001	0.98562329234424\\
0.0700000000000001	0.98283157788142\\
0.08	0.984138041321293\\
0.09	0.979714201620605\\
0.1	0.97729656293249\\
0.11	0.971615154962374\\
0.12	0.968154060311842\\
0.13	0.964113341386595\\
0.14	0.963144199666029\\
0.15	0.947797323896119\\
0.16	0.952793333194628\\
0.17	0.943794783060634\\
0.18	0.946714879696669\\
0.19	0.927090470469386\\
0.2	0.923686274509804\\
0.21	0.917098252334479\\
0.22	0.91121429630671\\
0.23	0.896024897439525\\
0.24	0.880133001171853\\
0.25	0.904883148684802\\
};
\addlegendentry{$k=$300}
\end{axis}
\end{tikzpicture}%

}
\caption{The robustness of our label mapping algorithm to mislabeled images for various numbers of feature clusters per class, $k$.}
\label{fig:robust_k}
\end{figure}
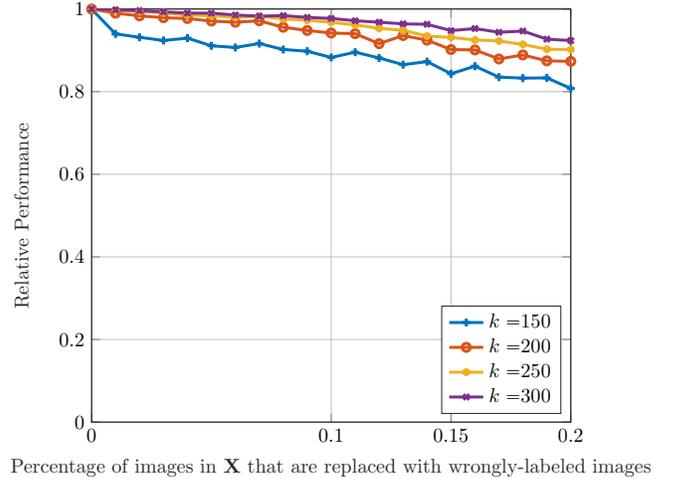

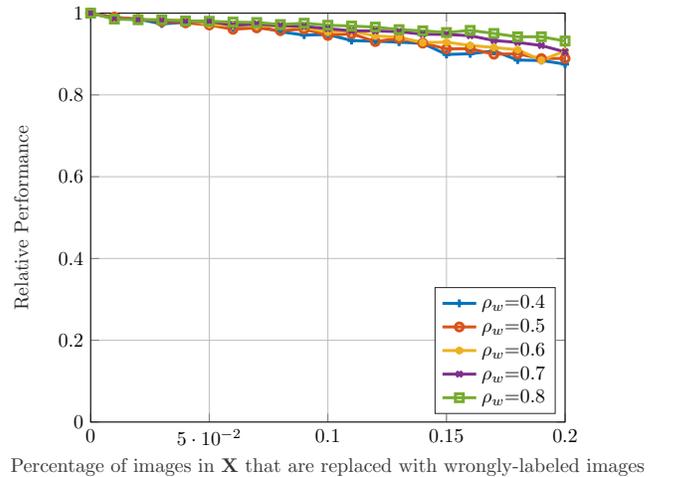
\begin{figure}[h]
\resizebox{\columnwidth}{!}{
\centering
\definecolor{mycolor1}{rgb}{0.00000,0.44700,0.74100}%
\definecolor{mycolor2}{rgb}{0.85000,0.32500,0.09800}%
\definecolor{mycolor3}{rgb}{0.92900,0.69400,0.12500}%
\definecolor{mycolor4}{rgb}{0.49400,0.18400,0.55600}%
\definecolor{mycolor5}{rgb}{0.46600,0.67400,0.18800}%

\begin{tikzpicture}

\begin{axis}[%
at={(1.525in,1in)},
scale only axis,
xmin=0,
xmax=0.2,
xlabel style={font=\color{white!20!black}},
xlabel={Percentage of images in $\mathbf{X}$ that are replaced with wrongly-labeled images},
xtick={0,0.05,0.1,0.15,0.2},
ymin=0,
ymax=1,
ylabel style={font=\color{white!15!black}},
ylabel={Relative Performance},
axis background/.style={fill=white},
xmajorgrids,
ymajorgrids,
legend style={at={(0.98,0.02)},anchor=south east,legend cell align=left, align=left, draw=white!15!black}
]
\addplot [color=mycolor1, line width=1.5pt, mark=+, mark options={solid, mycolor1}]
  table[row sep=crcr]{%
0	1\\
0.01	0.988609441641762\\
0.02	0.985464807847887\\
0.03	0.973587211622317\\
0.04	0.977920470693596\\
0.05	0.973718770519996\\
0.0600000000000001	0.964074366645726\\
0.0700000000000001	0.96838340385865\\
0.08	0.954510839431017\\
0.09	0.946373457187876\\
0.1	0.947659777429842\\
0.11	0.93302955241907\\
0.12	0.931359460230058\\
0.13	0.92908849570119\\
0.14	0.92595334742513\\
0.15	0.898972174648532\\
0.16	0.900629382041793\\
0.17	0.907775443639781\\
0.18	0.885347762379376\\
0.19	0.884137582169356\\
0.2	0.875392961886163\\
0.21	0.864709162430325\\
0.22	0.854426078135447\\
0.23	0.834188377493905\\
0.24	0.839625312330616\\
0.25	0.819590107093324\\
};
\addlegendentry{$\rho_w$=0.4}

\addplot [color=mycolor2, line width=1.5pt, mark=o, mark options={solid, mycolor2}]
  table[row sep=crcr]{%
0	1\\
0.01	0.990377990484665\\
0.02	0.985165077202912\\
0.03	0.979251055326394\\
0.04	0.976324712809618\\
0.05	0.971066138218133\\
0.0600000000000001	0.96055296416042\\
0.0700000000000001	0.963981706161315\\
0.08	0.957336899736699\\
0.09	0.9610087755384\\
0.1	0.945522905685516\\
0.11	0.949728185804928\\
0.12	0.930510277808681\\
0.13	0.938452765665487\\
0.14	0.926659534936096\\
0.15	0.912734253384495\\
0.16	0.913148764223342\\
0.17	0.899646395545464\\
0.18	0.900471075696045\\
0.19	0.888872412160791\\
0.2	0.889360885089043\\
0.21	0.87738081619813\\
0.22	0.862565254070649\\
0.23	0.858259570065525\\
0.24	0.843869794649367\\
0.25	0.828949626681814\\
};
\addlegendentry{$\rho_w$=0.5}

\addplot [color=mycolor3, line width=1.5pt, mark=asterisk, mark options={solid, mycolor3}]
  table[row sep=crcr]{%
0	1\\
0.01	0.988854613719478\\
0.02	0.984582936749914\\
0.03	0.98343794285688\\
0.04	0.982716387860232\\
0.05	0.972541406629162\\
0.0600000000000001	0.969363135496956\\
0.0700000000000001	0.968956249168608\\
0.08	0.96845137538056\\
0.09	0.96470652107875\\
0.1	0.953365280066411\\
0.11	0.956270004815318\\
0.12	0.944196687251111\\
0.13	0.941180220887595\\
0.14	0.928239302649419\\
0.15	0.929008531003959\\
0.16	0.921078155227681\\
0.17	0.915997411078669\\
0.18	0.91055326707331\\
0.19	0.884711829792933\\
0.2	0.907573506899954\\
0.21	0.893795210829181\\
0.22	0.876389225153414\\
0.23	0.874599876740696\\
0.24	0.856791691704636\\
0.25	0.841512774406741\\
};
\addlegendentry{$\rho_w$=0.6}

\addplot [color=mycolor4, line width=1.5pt, mark=x, mark options={solid, mycolor4}]
  table[row sep=crcr]{%
0	1\\
0.01	0.98792331760691\\
0.02	0.984233369323101\\
0.03	0.981640057292176\\
0.04	0.979250266487161\\
0.05	0.980230975955424\\
0.0600000000000001	0.970267160249981\\
0.0700000000000001	0.973495396546158\\
0.08	0.969468603361966\\
0.09	0.967595339367265\\
0.1	0.962038666303277\\
0.11	0.956911179553889\\
0.12	0.956475192264873\\
0.13	0.954757315240576\\
0.14	0.948403457338538\\
0.15	0.948001100998387\\
0.16	0.94481319649015\\
0.17	0.933092090439181\\
0.18	0.928471916724357\\
0.19	0.920831881153612\\
0.2	0.905217716215775\\
0.21	0.905113027627898\\
0.22	0.905428748022062\\
0.23	0.885839746520988\\
0.24	0.870987668092654\\
0.25	0.867446891020704\\
};
\addlegendentry{$\rho_w$=0.7}

\addplot [color=mycolor5, line width=1.5pt, mark=square, mark options={solid, mycolor5}]
  table[row sep=crcr]{%
0	1\\
0.01	0.985503347367642\\
0.02	0.983857277189811\\
0.03	0.983809936329042\\
0.04	0.980761566890268\\
0.05	0.980132323927506\\
0.0600000000000001	0.978160202243948\\
0.0700000000000001	0.977194420464216\\
0.08	0.972228865972165\\
0.09	0.975416435652069\\
0.1	0.970526694677212\\
0.11	0.968467988571786\\
0.12	0.965915364209923\\
0.13	0.959751473803228\\
0.14	0.956739952387621\\
0.15	0.952673754899836\\
0.16	0.957875542121172\\
0.17	0.950397133379689\\
0.18	0.942142398357763\\
0.19	0.942272914056016\\
0.2	0.93189345870586\\
0.21	0.922867501989711\\
0.22	0.931583514985399\\
0.23	0.923424647897232\\
0.24	0.921919196119948\\
0.25	0.913771993038162\\
};
\addlegendentry{$\rho_w$=0.8}

\end{axis}
\end{tikzpicture}

}
\caption{The robustness of our label mapping algorithm to mislabeled images for various feature sparsity levels, $\rho_w$.}
\label{fig:robust_sw}
\end{figure}

Since our similarity-based retrieval workflow might produce a few images that do not belong to the same class as the reference image, we might end up with a few wrong image-level labels. However, the $k$-means initialization step of $\mathbf{W}^0$ greatly enhances the robustness of our label mapping algorithm to mislabeled images. To validate this claim, we examine the effect of wrongly retrieved images on the final pixel-level labels and analyze the robustness of our label mapping algorithm. We achieve this by artificially replacing images in $\mathbf{X}$ with wrongly-labeled images, and then computing the final pixel-level labels and comparing the performance of our label mapping algorithm relative to the base case where no images are replaced. The performance is evaluated using a metric called \textit{pixel accuracy} that measures the percentage of all pixels that are correctly classified. Pixels with low confidence in the base case are ignored. Fig. \ref{fig:robust_k} shows the drop in relative performance as the percentage of wrongly-labeled images in $\mathbf{X}$ increases for varying numbers of feature clusters per class, $k$. Fig. \ref{fig:robust_sw} shows a similar drop for different values of the feature sparsity $\rho_w$. Overall, the larger the number of clusters, and the higher the sparsity of the initial features, the more robust the label mapping algorithm is to wrongly retrieved images. 


We then apply the MURs in equations \ref{WMUR} and \ref{HMUR} successively until both $\mathbf{W}$ and $\mathbf{H}$ converge. Figure \ref{fig:optimization} shows the convergence curves for the $\mathbf{W}$ objective function in equation \ref{eq:WObjective}, the $\mathbf{H}$ objective function in equation \ref{eq:HObjective}, and the overall objective function defined in equation \ref{SONMF}. We see that although we did not attempt to solve equation \ref{SONMF} directly, the two MURs in equations \ref{WMUR} and \ref{HMUR} effectively minimize the overall objective function. We terminate our optimization at the $200^\text{th}$ iteration. Performed on a GPU, this process takes around 4 seconds. 
\input{other/optimization_figure}

\begin{figure*}[]
\begin{center}
\begin{tabular}{c|c|c|c|c|c}
\includegraphics[width=2.2cm]{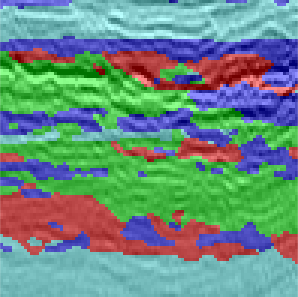} &
\includegraphics[width=2.2cm]{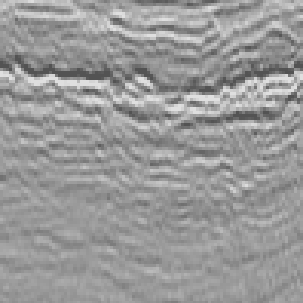} &
\includegraphics[width=2.2cm]{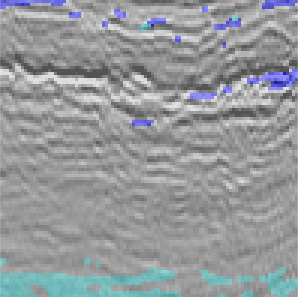} &
\includegraphics[width=2.2cm]{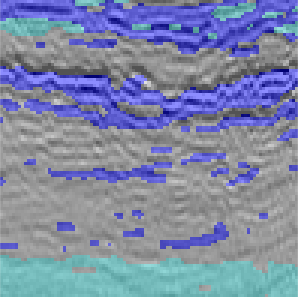} &
\includegraphics[width=2.2cm]{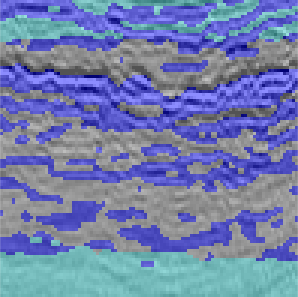} &
\includegraphics[width=2.2cm]{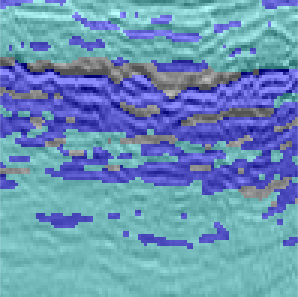} \\ 
\includegraphics[width=2.2cm]{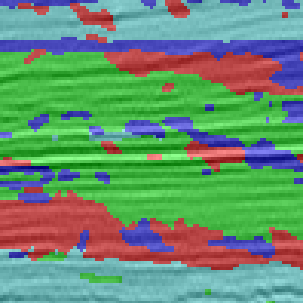} &
\includegraphics[width=2.2cm]{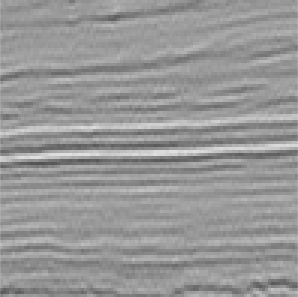} &
\includegraphics[width=2.2cm]{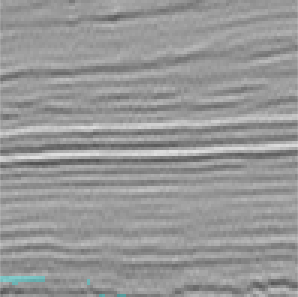}&
\includegraphics[width=2.2cm]{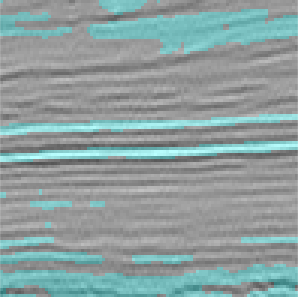} &
\includegraphics[width=2.2cm]{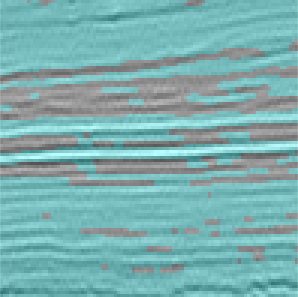} &
\includegraphics[width=2.2cm]{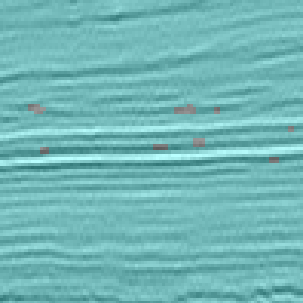} \\ 
\includegraphics[width=2.2cm]{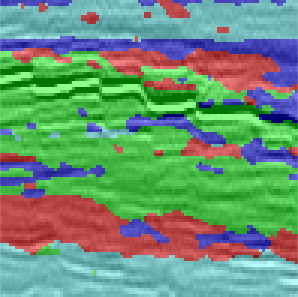} &
\includegraphics[width=2.2cm]{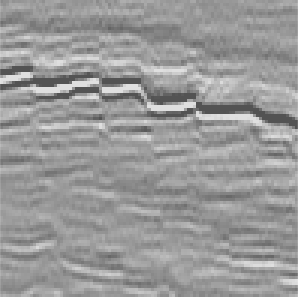} &
\includegraphics[width=2.2cm]{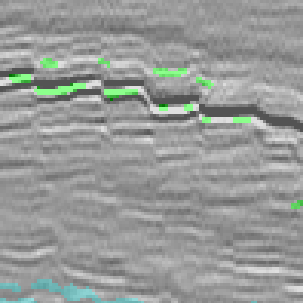} &
\includegraphics[width=2.2cm]{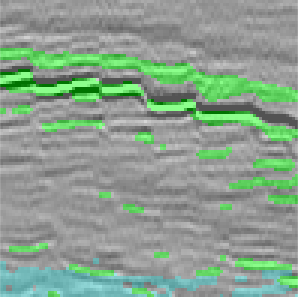} &
\includegraphics[width=2.2cm]{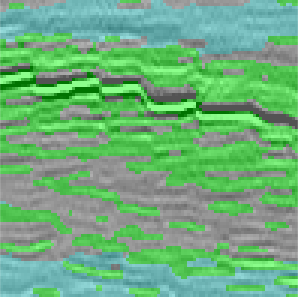} &
\includegraphics[width=2.2cm]{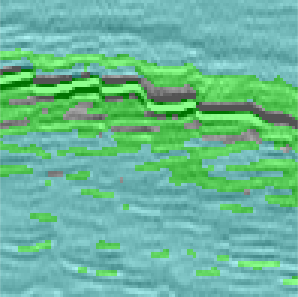} \\ 

\includegraphics[width=2.2cm]{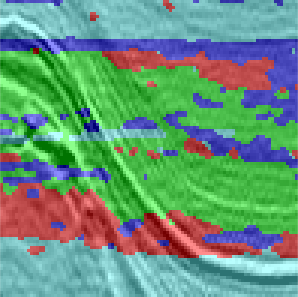} &
\includegraphics[width=2.2cm]{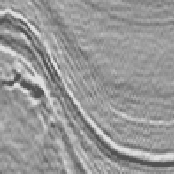} &
\includegraphics[width=2.2cm]{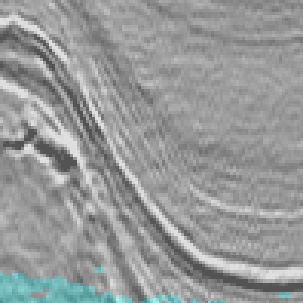} &
\includegraphics[width=2.2cm]{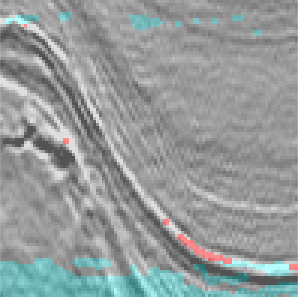} &
\includegraphics[width=2.2cm]{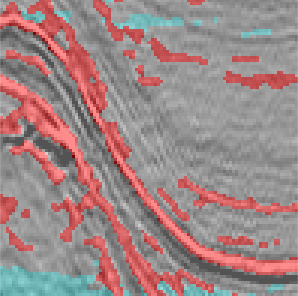} &
\includegraphics[width=2.2cm]{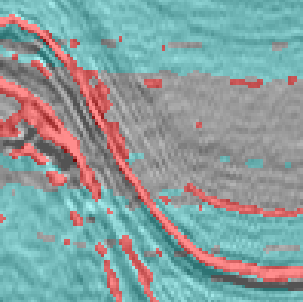} \\
Initialization & $1^\text{st}$ iteration  & $10^\text{{th}}$ iteration & $20^\text{{th}}$ iteration & $50^\text{{th}}$ iteration & $200^\text{{th}}$ iteration 
\end{tabular}
\end{center}

\captionof{figure}{Results of our weakly supervised label mapping approach for sample images from each class for various iterations. The initial labels (i.e. with randomly initialized coefficients) are also shown in the first column.}
\label{fig:evolution}
\end{figure*}

Figure \ref{fig:evolution} shows the initial labels for four different images from the four different classes, as well as the labels for various iterations in the optimization process. We note that since the coefficient matrix $\mathbf{H}$ was initialized with random values in the range $[0,1]$, our ``initial'' confidence computed using equation \ref{eqn:confidence} is very high, and consequently, very few pixels in the initial labels had coefficient smaller than $\tau$ and therefore belonged to the \texttt{uncertain} class). However, immediately after we start applying the MURs the confidence value of the labels drastically drops, to the degree that all the pixels in images during the first iteration are labeled as \texttt{uncertain}. However, as the optimization progresses, confidence in various predicted labels gradually increases. Towards the end of the optimization, the orthogonality term in equation \ref{SONMF} plays a more prominent role in ensuring that most features in $\mathbf{W}$ represent only a few images in $\mathbf{X}$, this significantly reduces the number of noisy labels.   

Figure \ref{fig:results} displays several examples selected at random from the final results that we obtain for various classes of subsurface structures. These results are shown for the baseline case using NMF only, the case where we use sparse initial features in $\mathbf{W}^0$, and our proposed formulation. We observe that the NMF results are very noisy. The results with sparse initial features are better, but they contain bands of misclassified pixels, typically in the center of the image. However, the results of our proposed formulation are much better, and they do not exhibit the same misclassified bands. Our results show a very good match between the labeled subsurface structures and the structures in the original seismic image. Furthermore, it is important to note that this method is not limited to these particular classes of subsurface structures and can be easily applied to any other structure as long as a sufficient number of similar images are retrieved for each class.  
\newcommand{\STAB}[1]{\begin{tabular}{@{}c@{}}#1\end{tabular}}

\begin{figure*}[]
  \centering
\begin{center}
\begin{tabular}{cc|c|c|c}
&{\color{blue}\texttt{chaotic}} images & {\color{cyan}\texttt{other}} images & {\color{green}\texttt{fault}} images & {\color{red}\texttt{salt dome}} images
\\
\multirow{3}{*}{\STAB{\rotatebox[origin=c]{90}{\footnotesize NMF~~~~~}}}
&
\includegraphics[width=1.55cm]{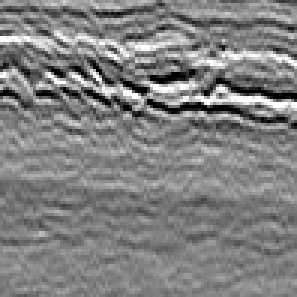} 
\includegraphics[width=1.55cm]{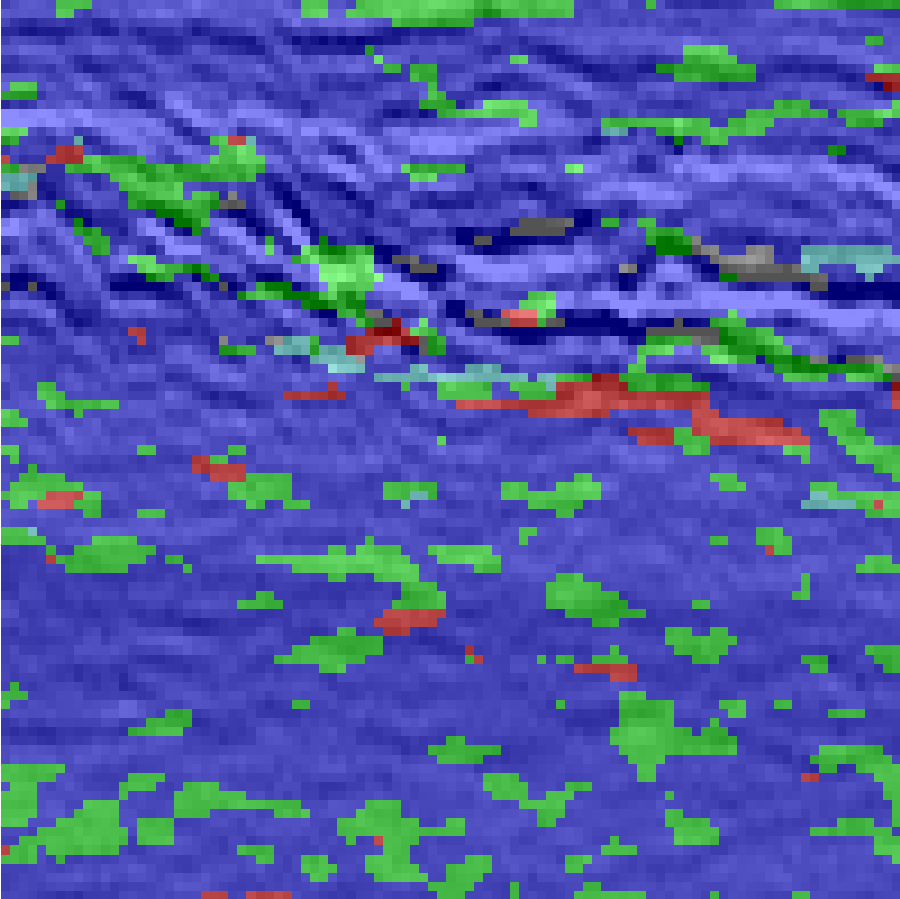}
&
\includegraphics[width=1.55cm]{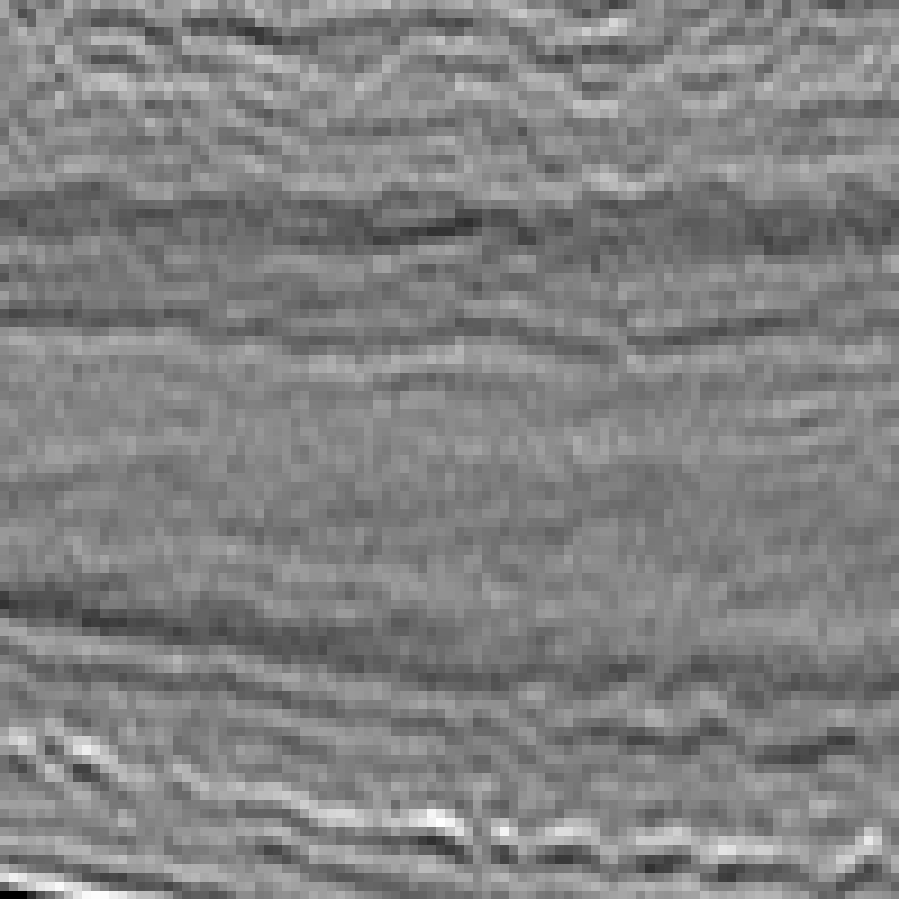} 
\includegraphics[width=1.55cm]{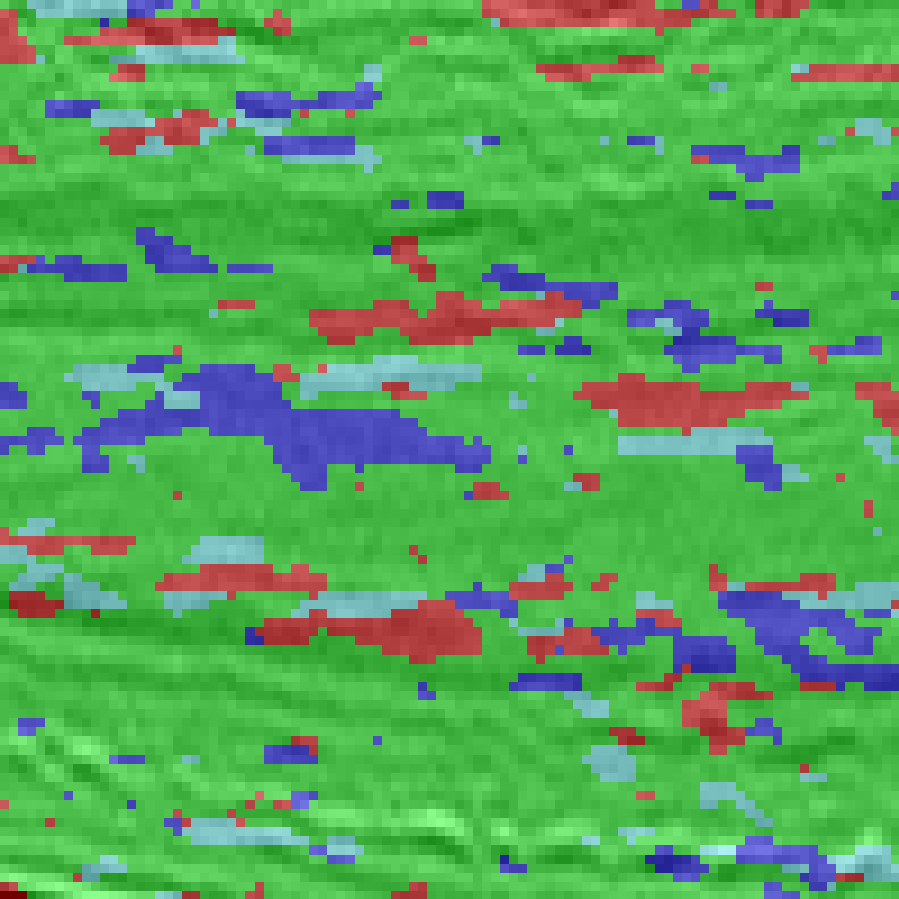}
&
\includegraphics[width=1.55cm]{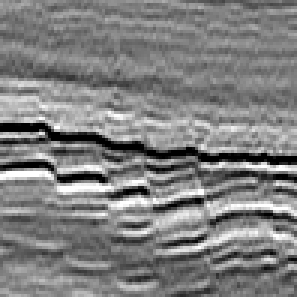} 
\includegraphics[width=1.55cm]{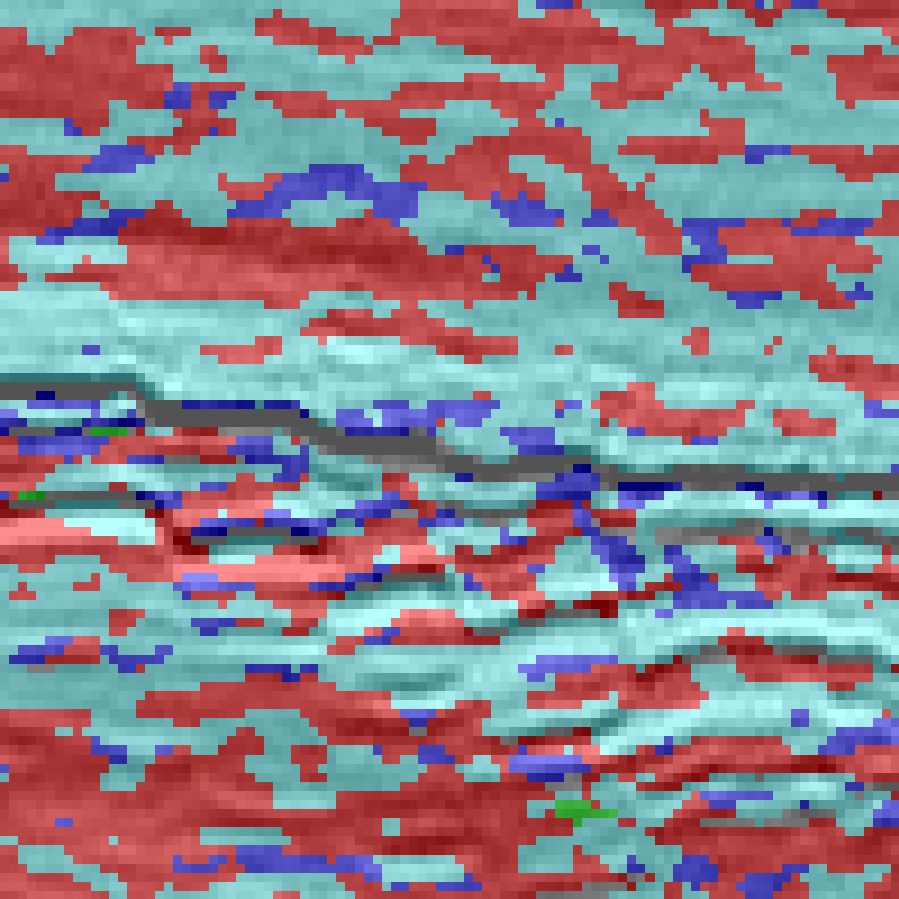}
&
\includegraphics[width=1.55cm]{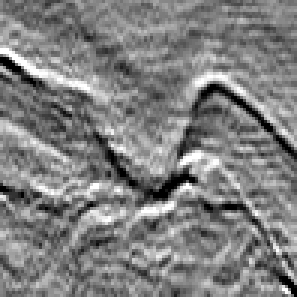} 
\includegraphics[width=1.55cm]{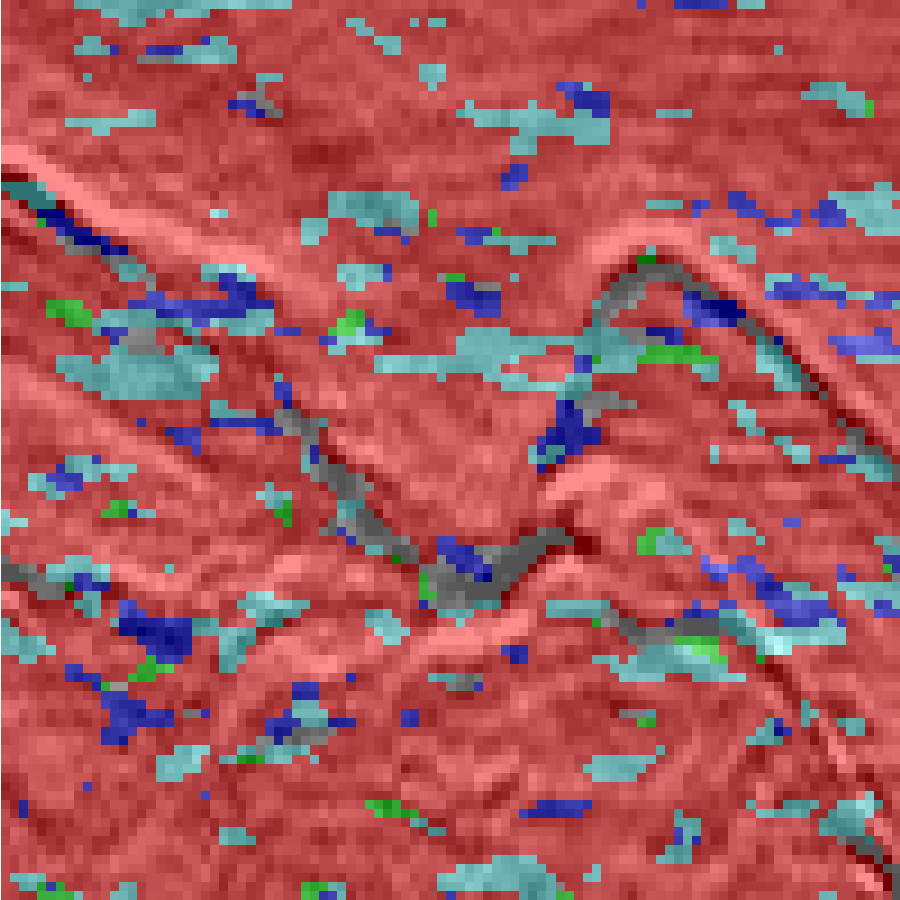}
\\
&
\includegraphics[width=1.55cm]{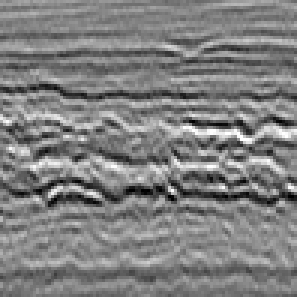} 
\includegraphics[width=1.55cm]{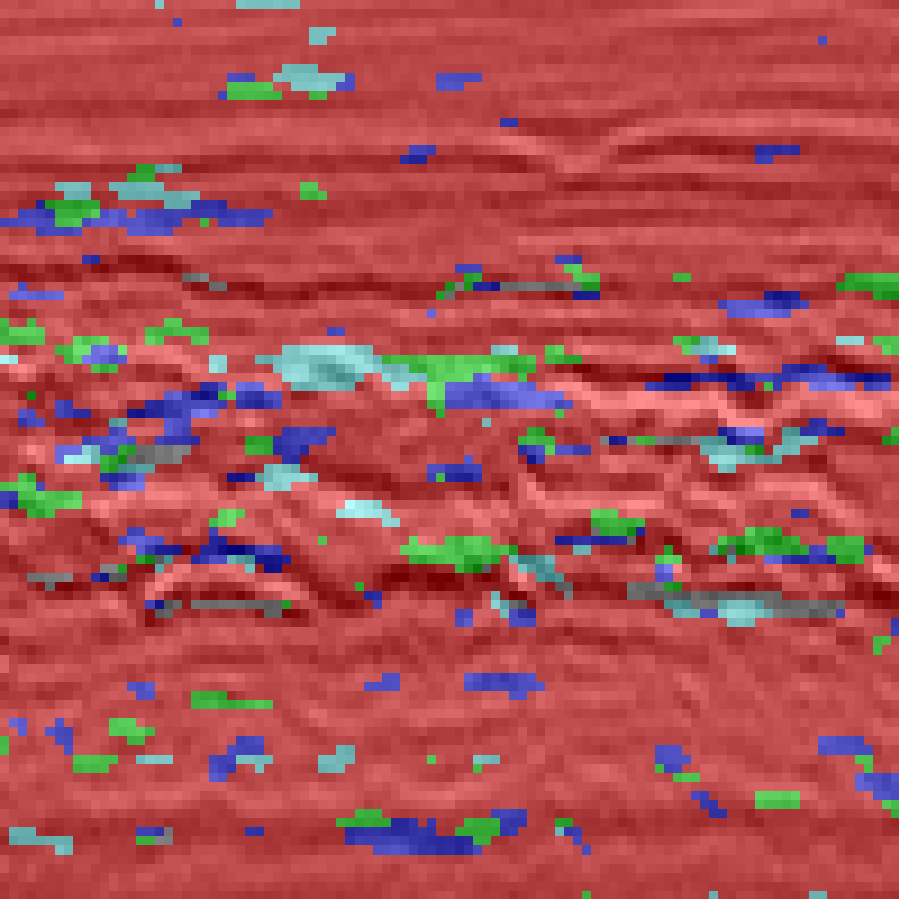}
&
\includegraphics[width=1.55cm]{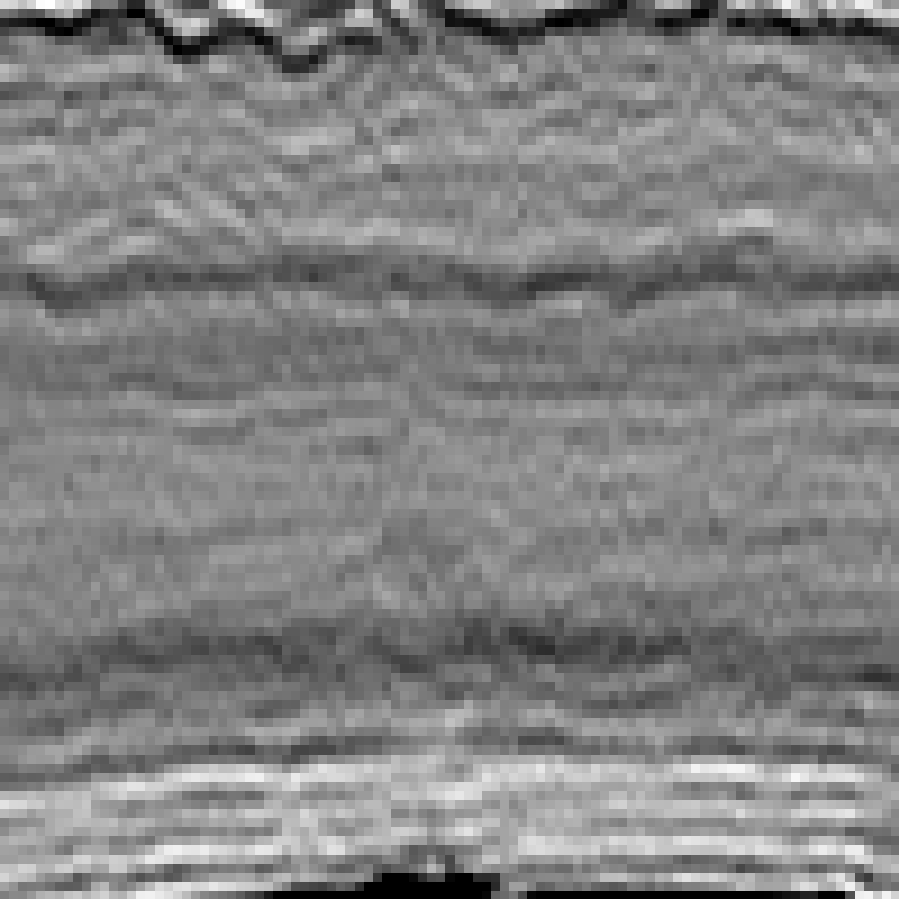} 
\includegraphics[width=1.55cm]{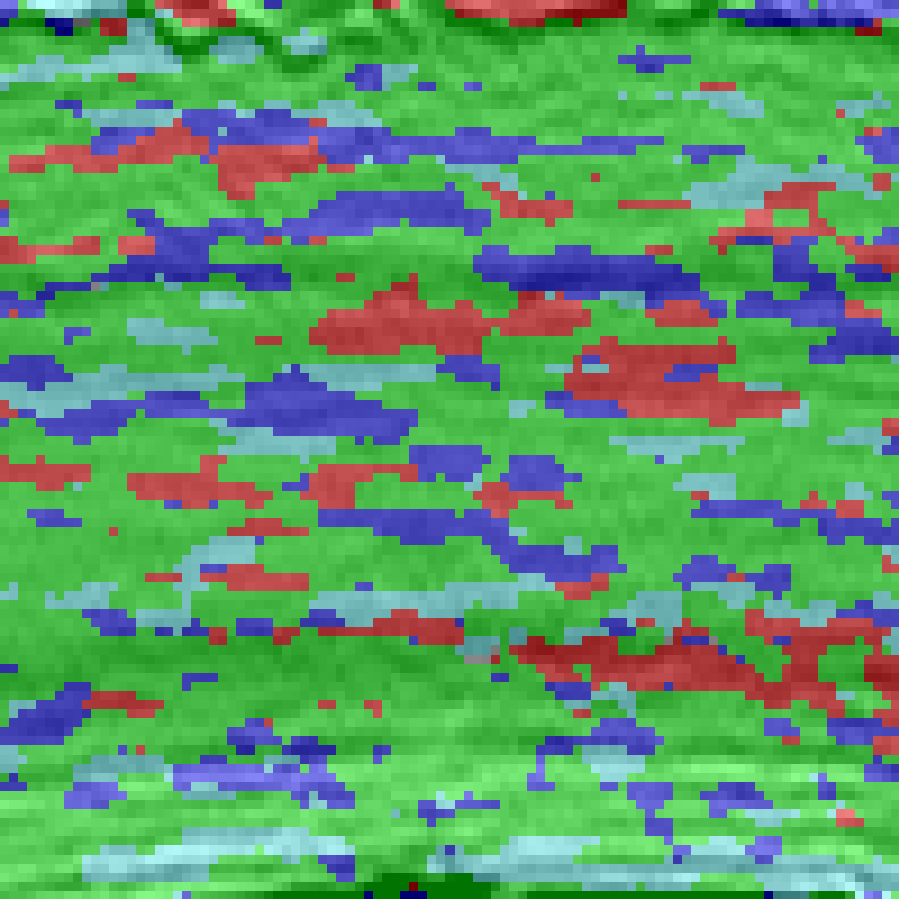}
&
\includegraphics[width=1.55cm]{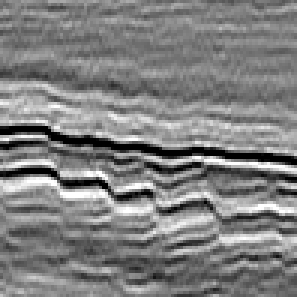} 
\includegraphics[width=1.55cm]{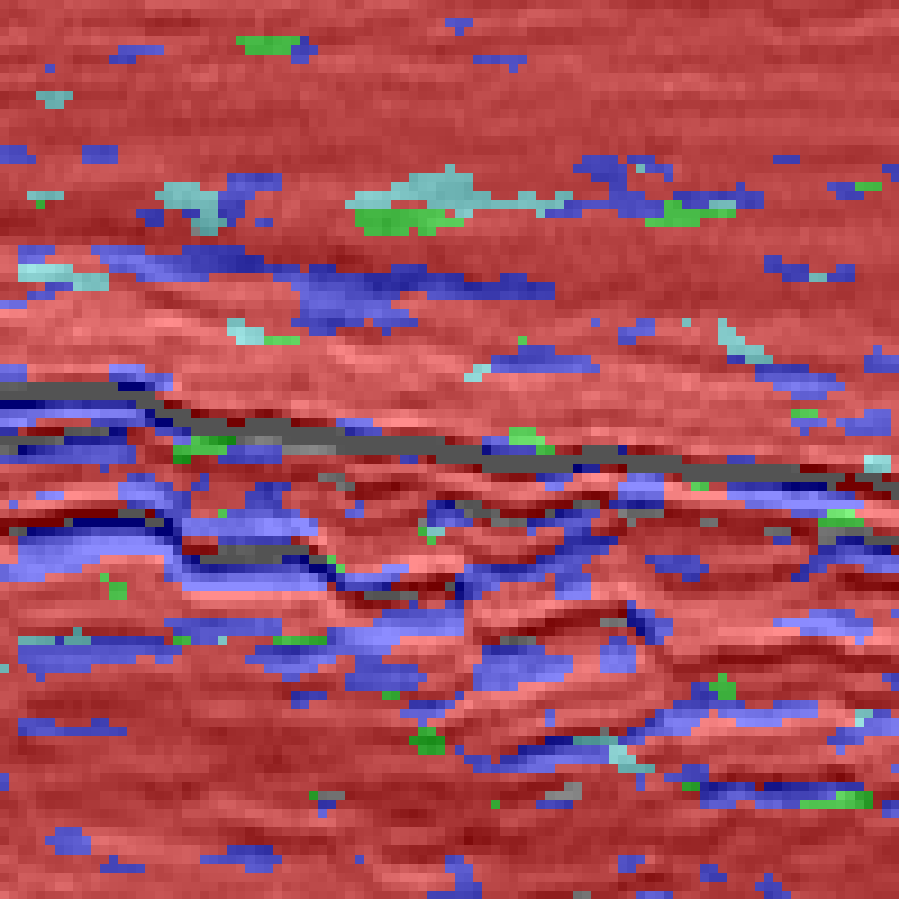}
&
\includegraphics[width=1.55cm]{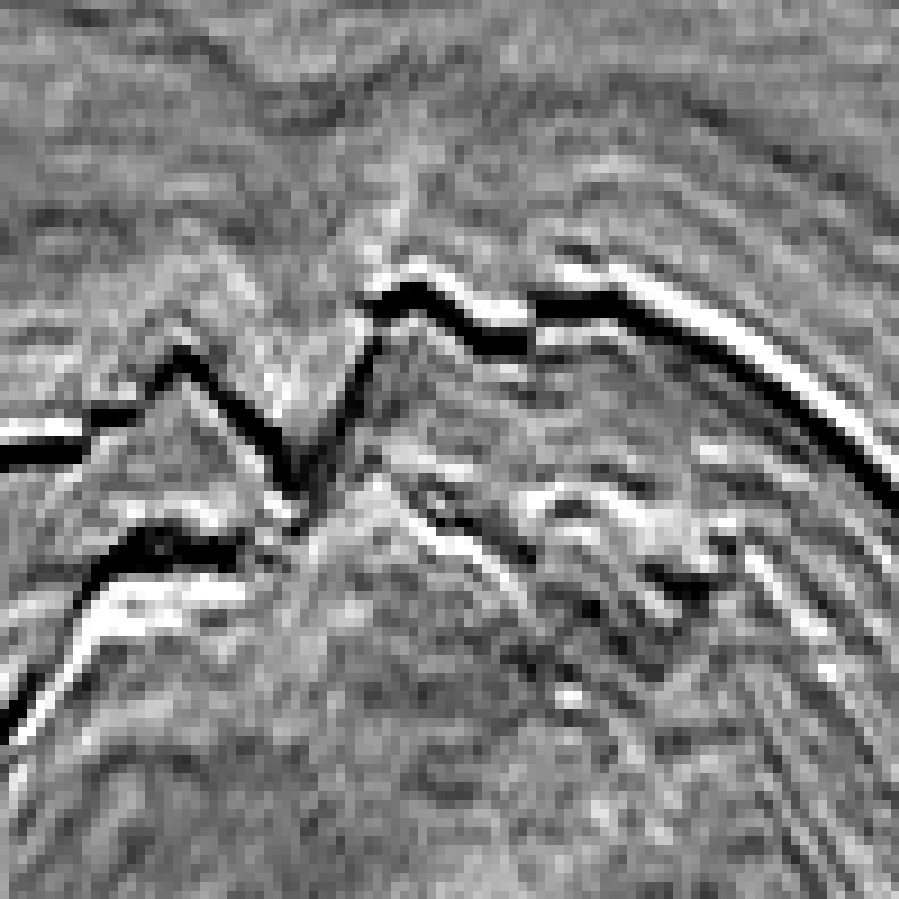} 
\includegraphics[width=1.55cm]{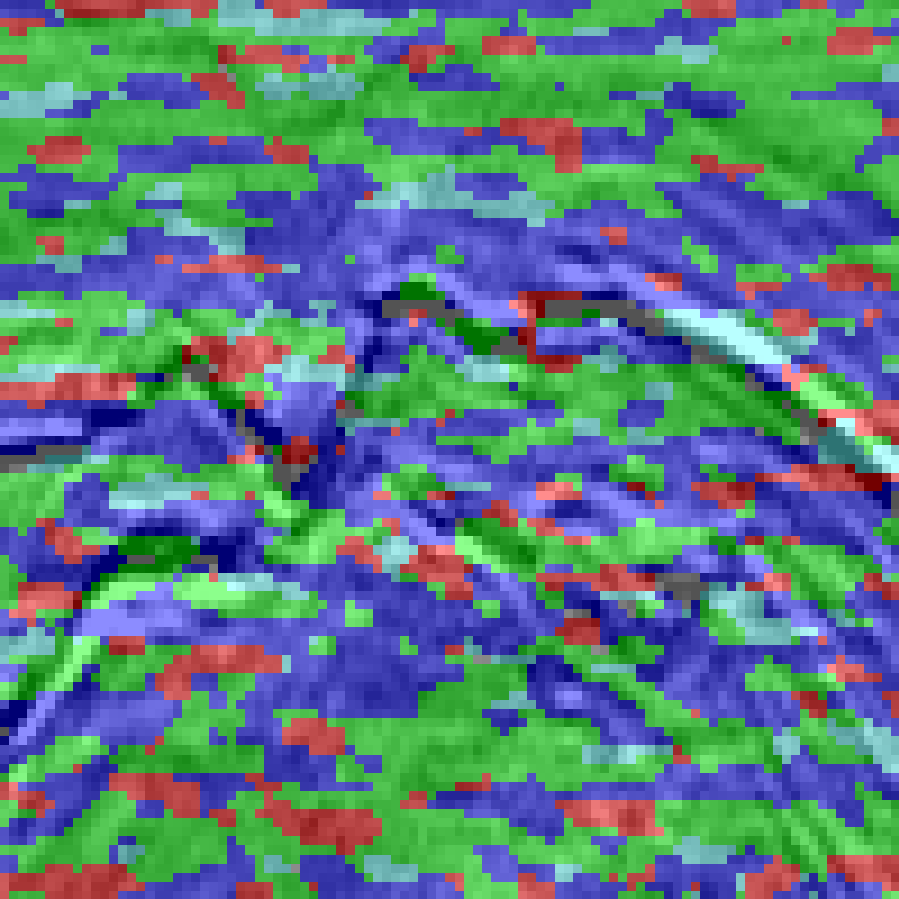}
\\
&
\includegraphics[width=1.55cm]{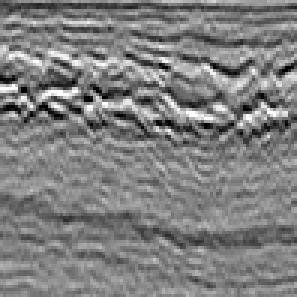} 
\includegraphics[width=1.55cm]{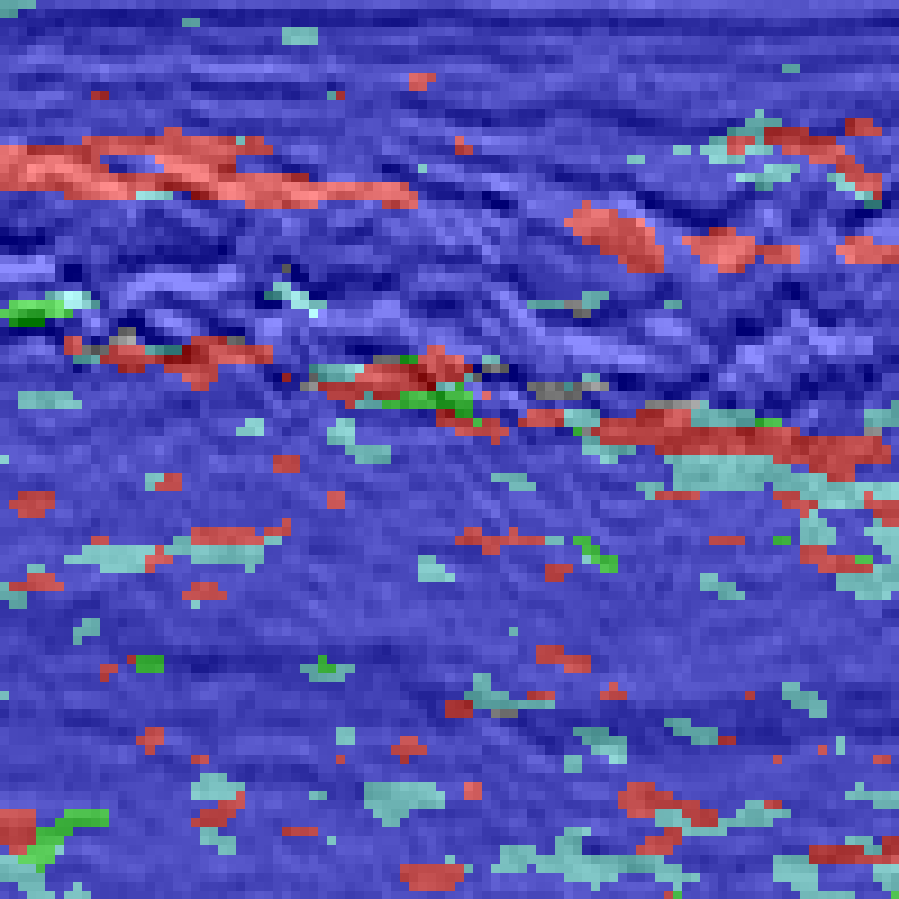}
&
\includegraphics[width=1.55cm]{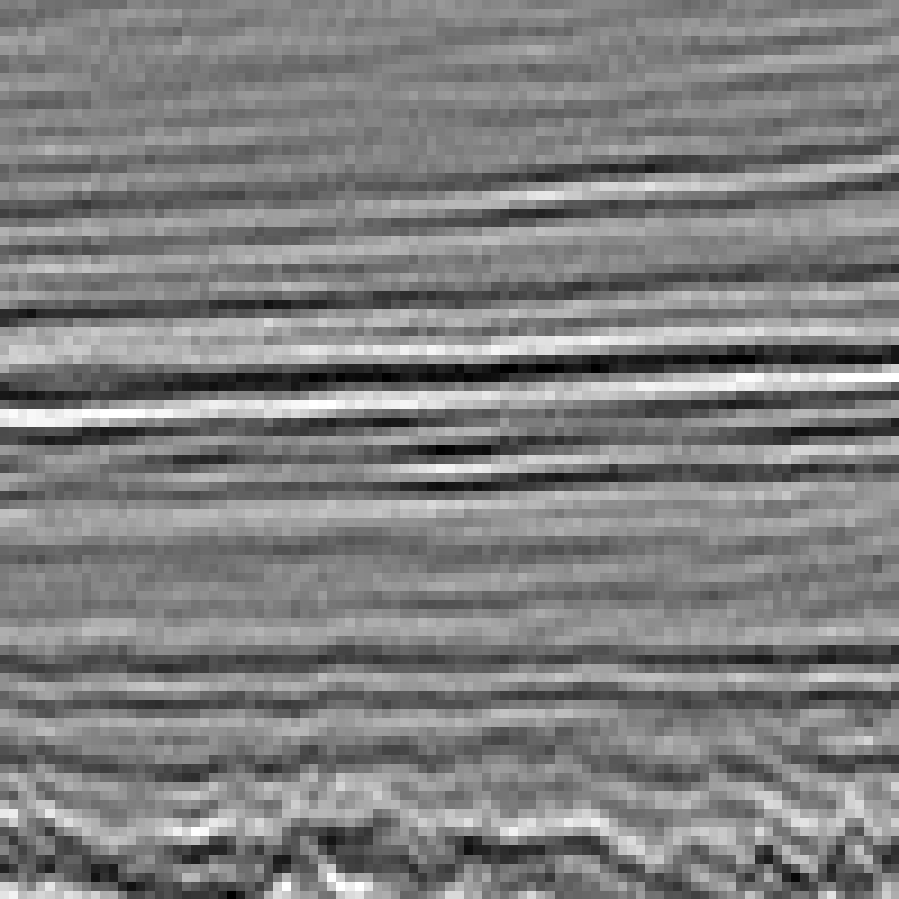} 
\includegraphics[width=1.55cm]{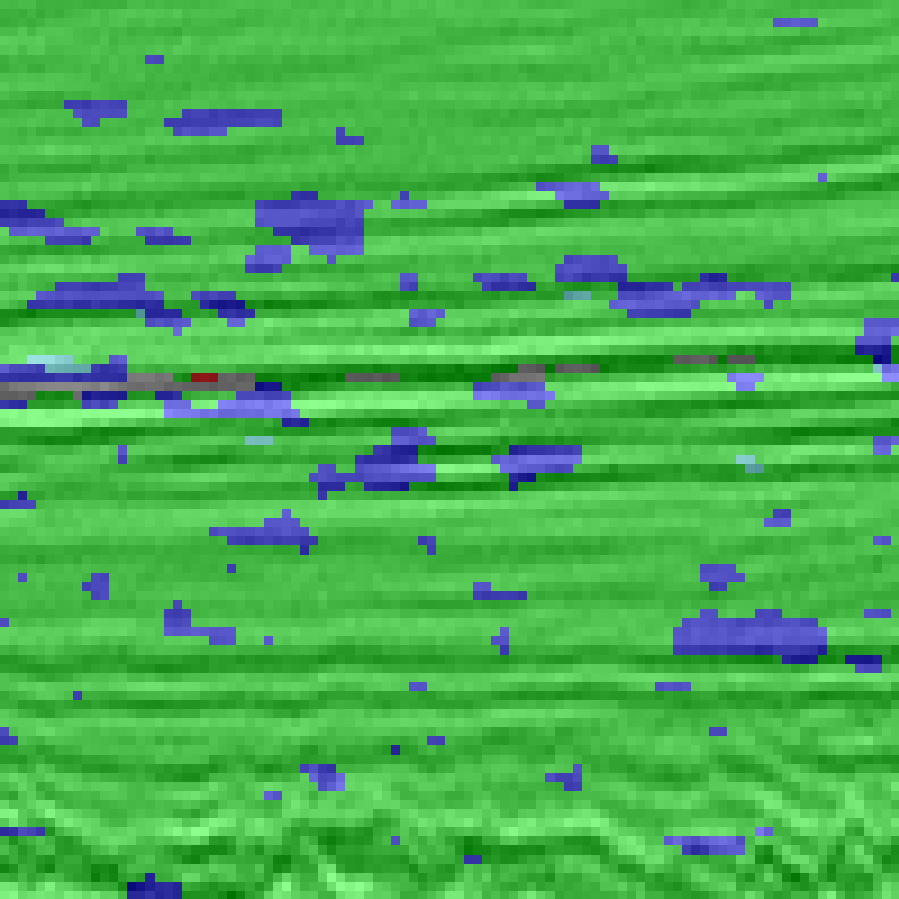}
&
\includegraphics[width=1.55cm]{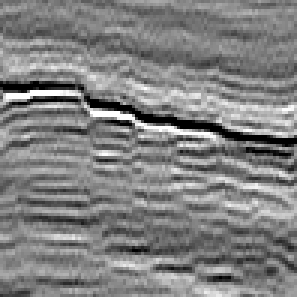} 
\includegraphics[width=1.55cm]{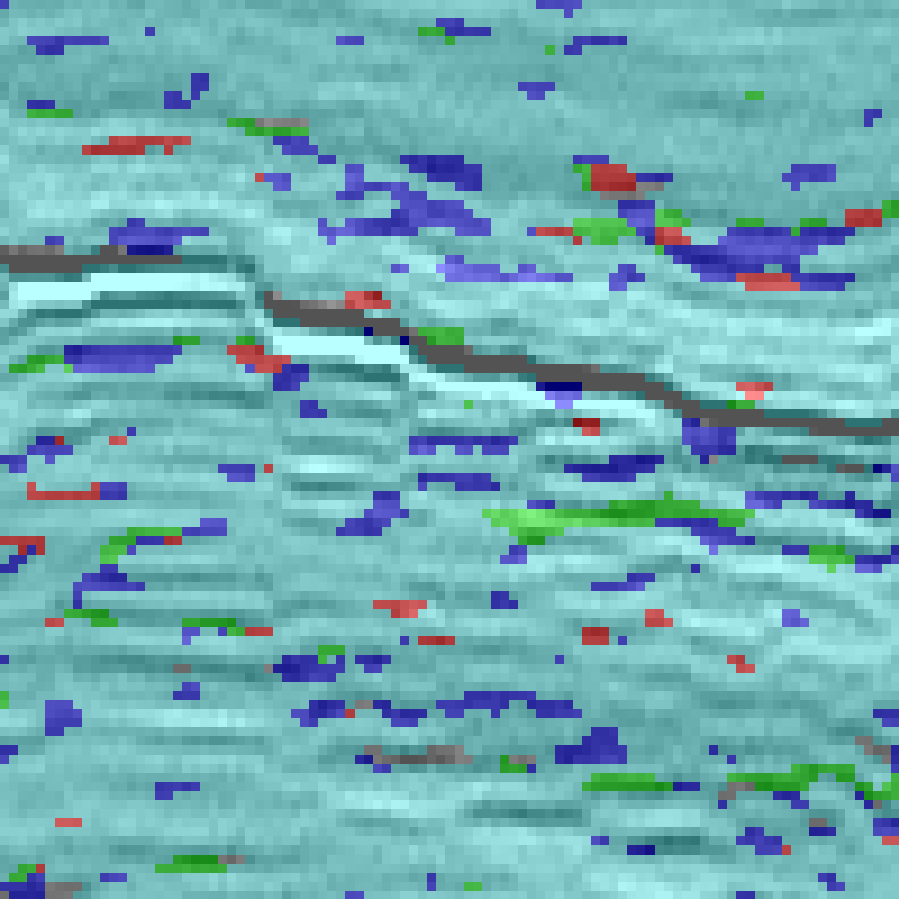}
&
\includegraphics[width=1.55cm]{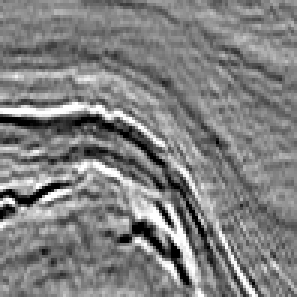} 
\includegraphics[width=1.55cm]{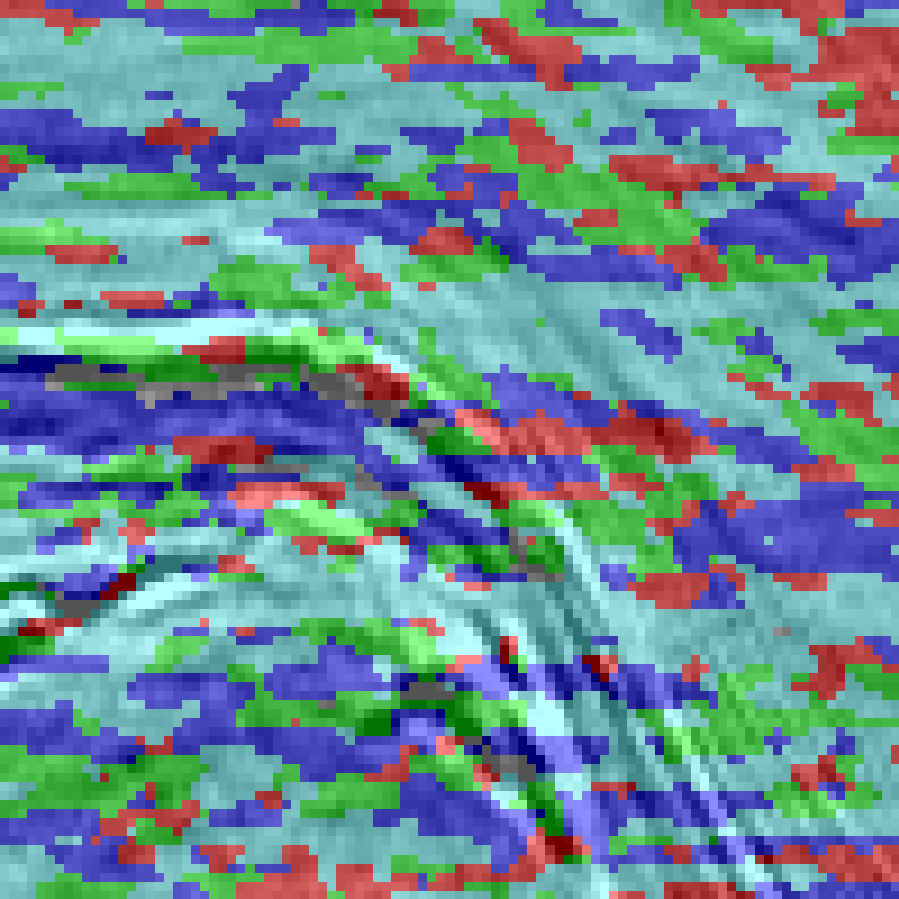}
\\
& & & 
\\
\multirow{3}{*}{\STAB{\rotatebox[origin=c]{90}{\footnotesize NMF with sparse features}}}
&
\includegraphics[width=1.55cm]{figures/images/img_1.png} 
\includegraphics[width=1.55cm]{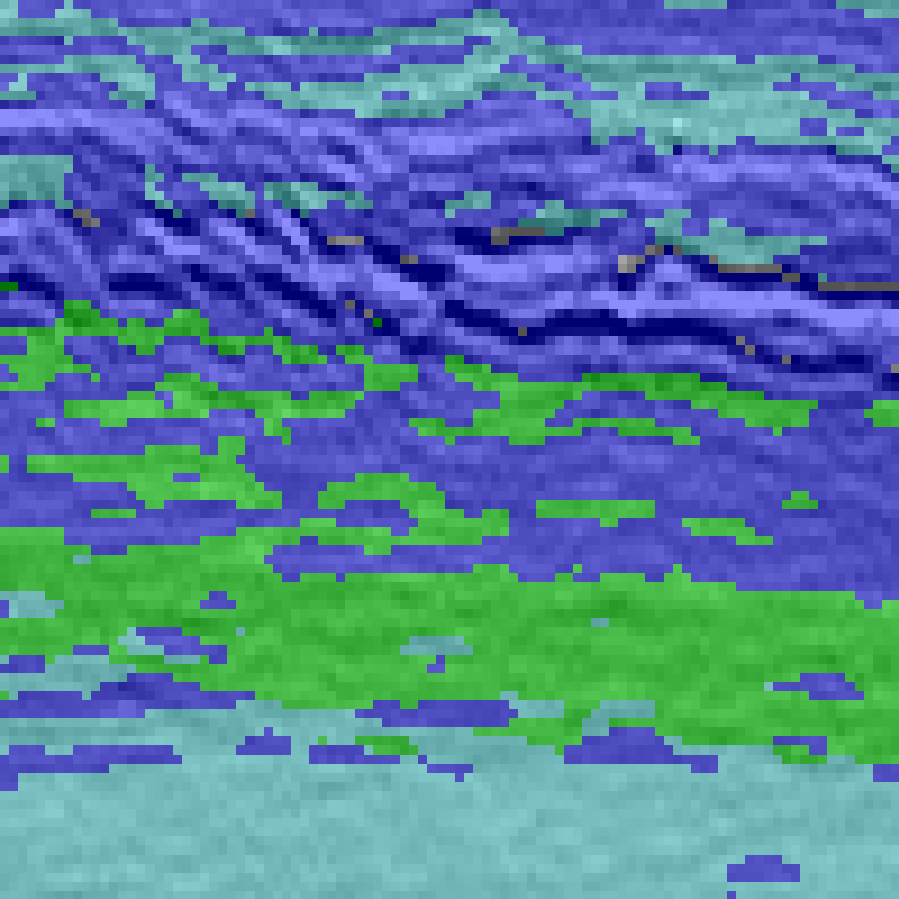}
&
\includegraphics[width=1.55cm]{figures/images/img_520.png} 
\includegraphics[width=1.55cm]{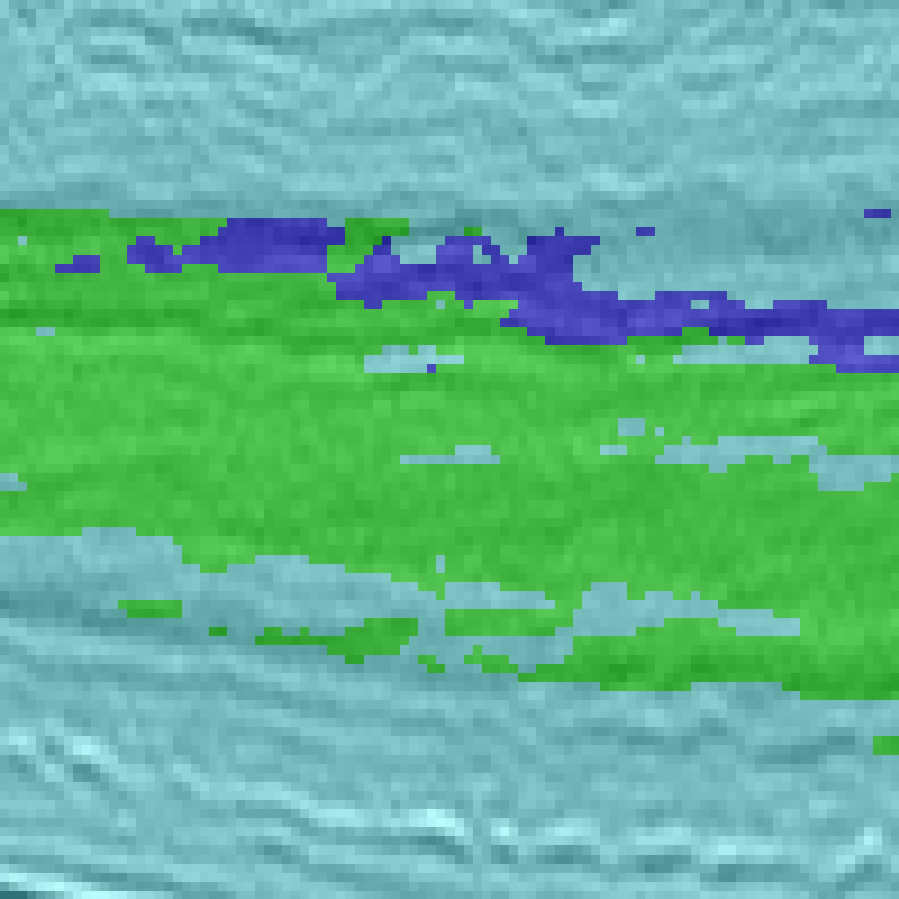}
&
\includegraphics[width=1.55cm]{figures/images/img_1026.png} 
\includegraphics[width=1.55cm]{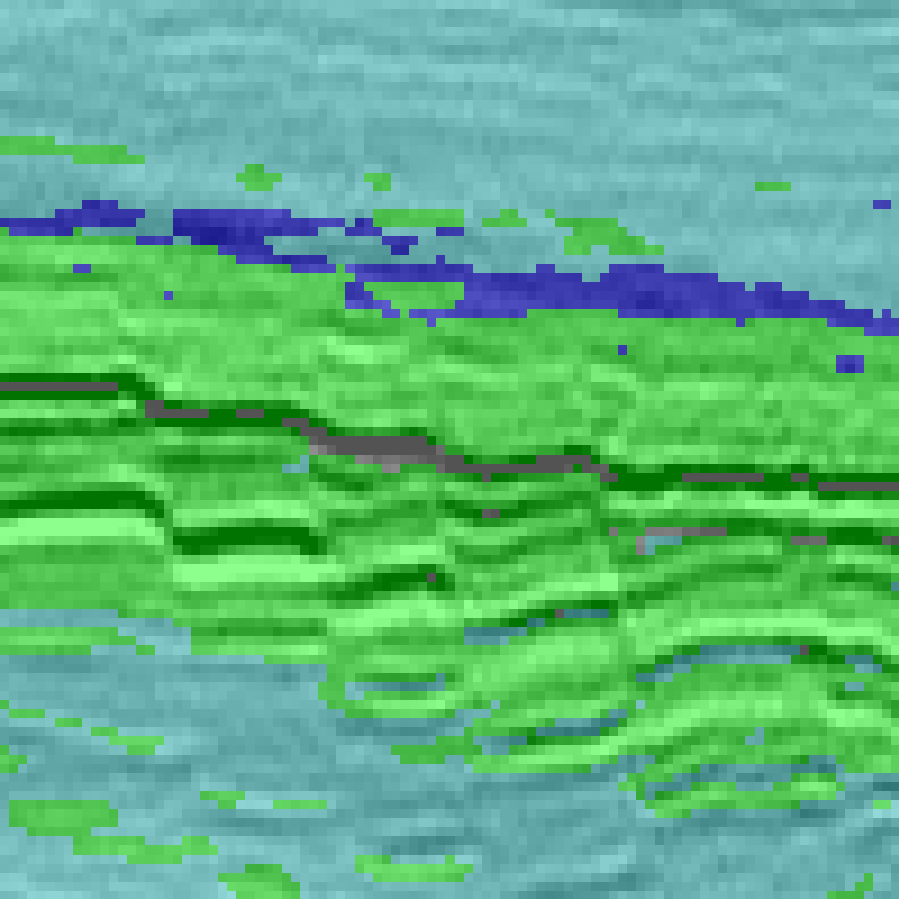}
&
\includegraphics[width=1.55cm]{figures/images/img_1539.png} 
\includegraphics[width=1.55cm]{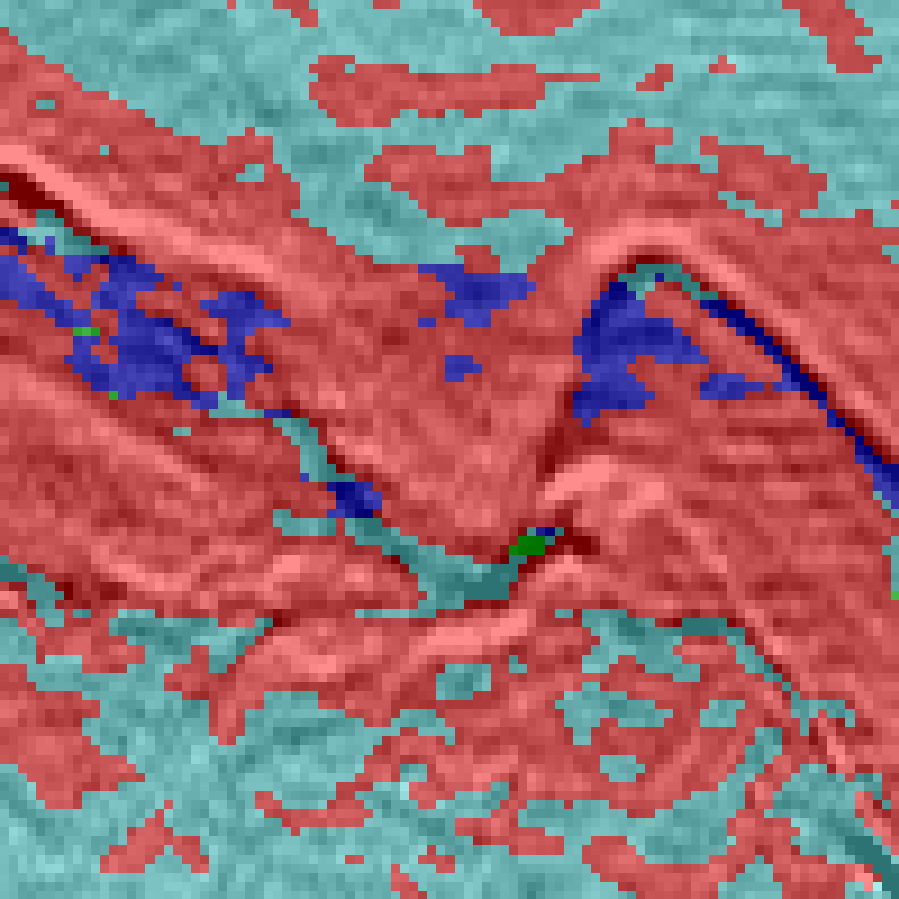}
\\
&
\includegraphics[width=1.55cm]{figures/images/img_104.png} 
\includegraphics[width=1.55cm]{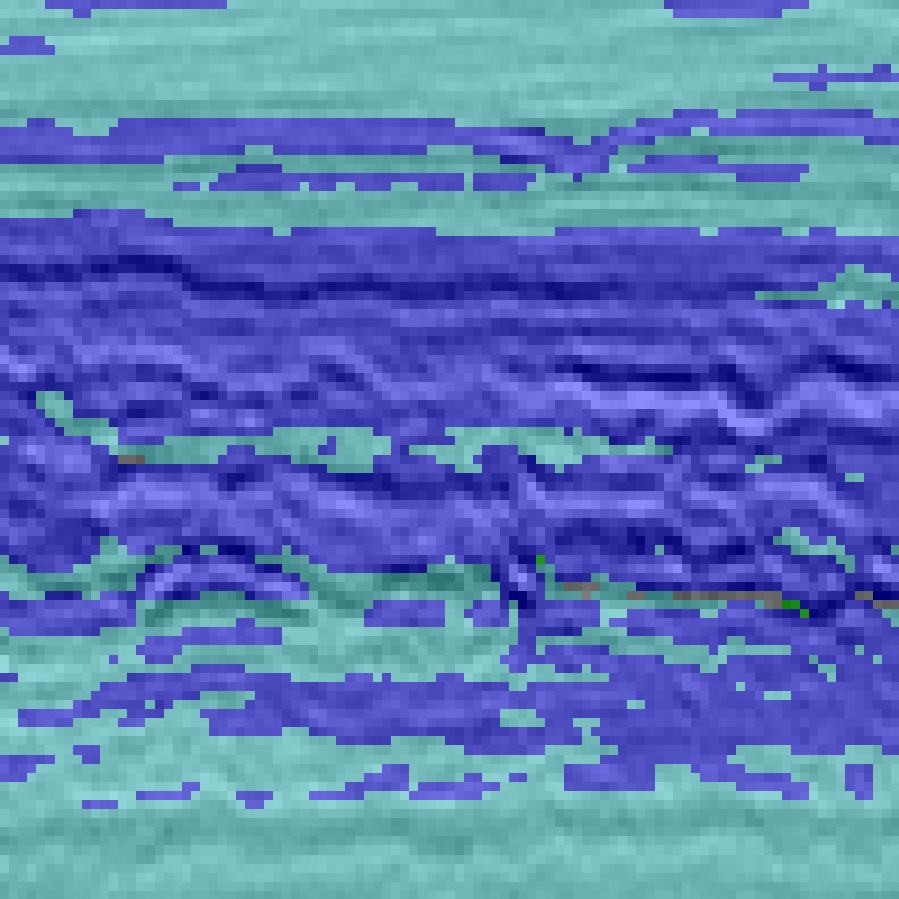}
&
\includegraphics[width=1.55cm]{figures/images/img_527.png} 
\includegraphics[width=1.55cm]{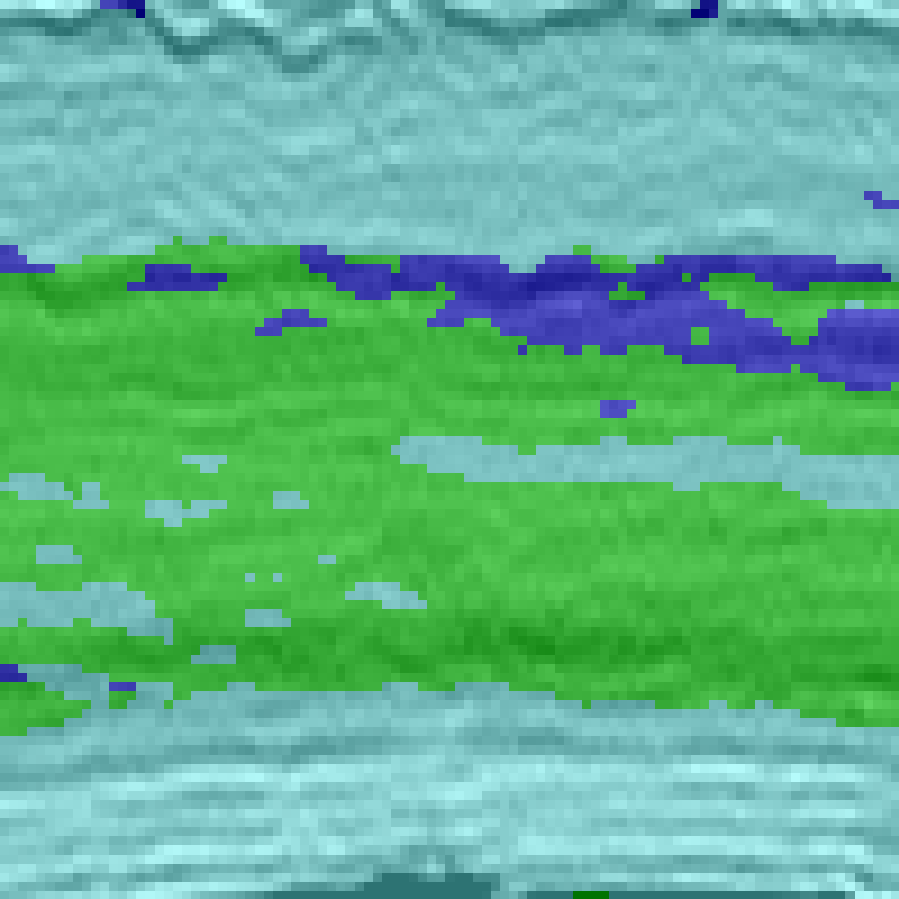}
&
\includegraphics[width=1.55cm]{figures/images/img_1129.png}
\includegraphics[width=1.55cm]{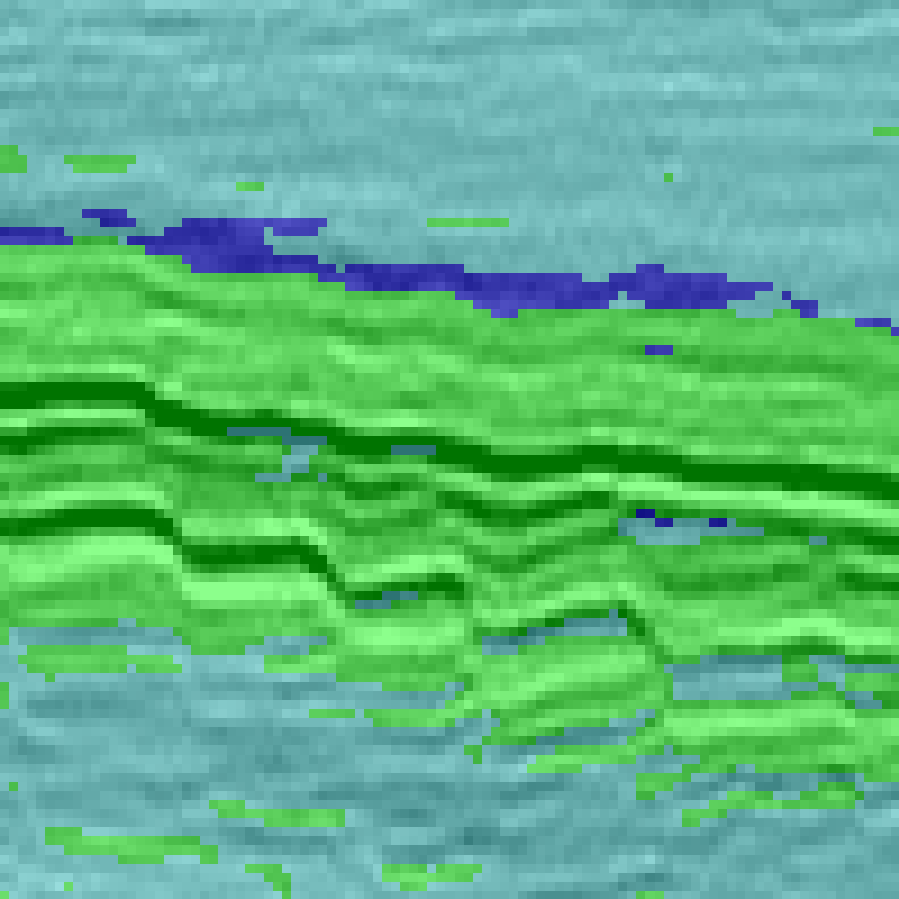}
&
\includegraphics[width=1.55cm]{figures/images/img_1638.png}
\includegraphics[width=1.55cm]{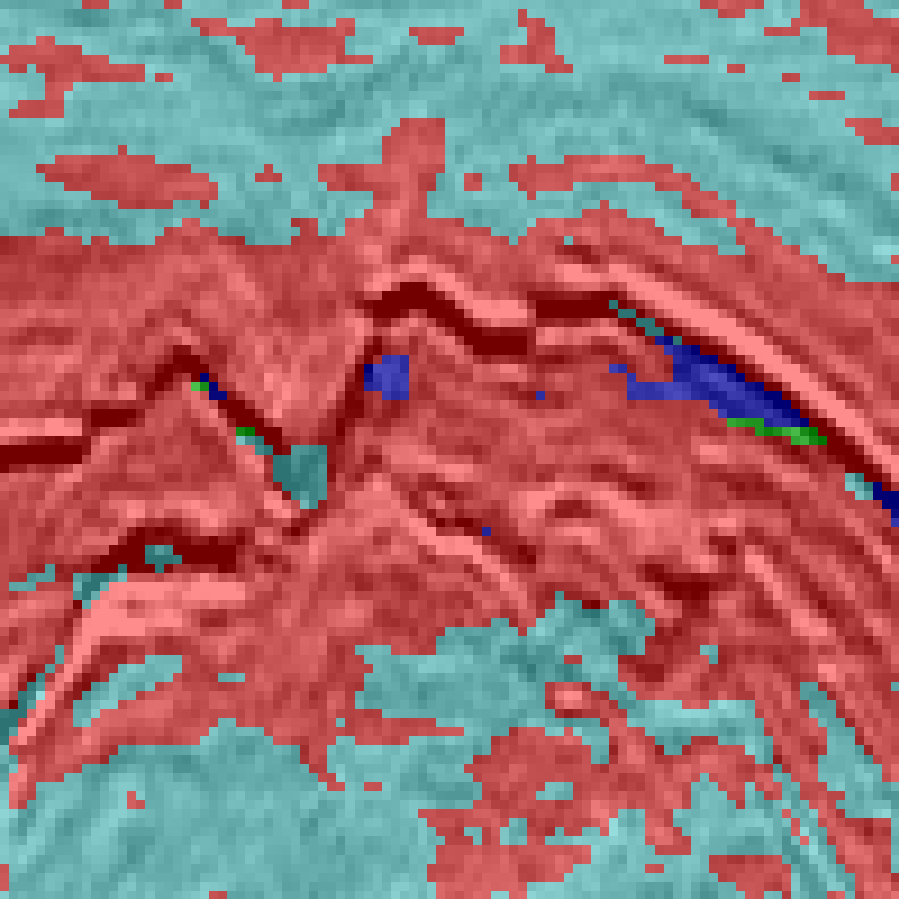}
\\
&
\includegraphics[width=1.55cm]{figures/images/img_360.png} 
\includegraphics[width=1.55cm]{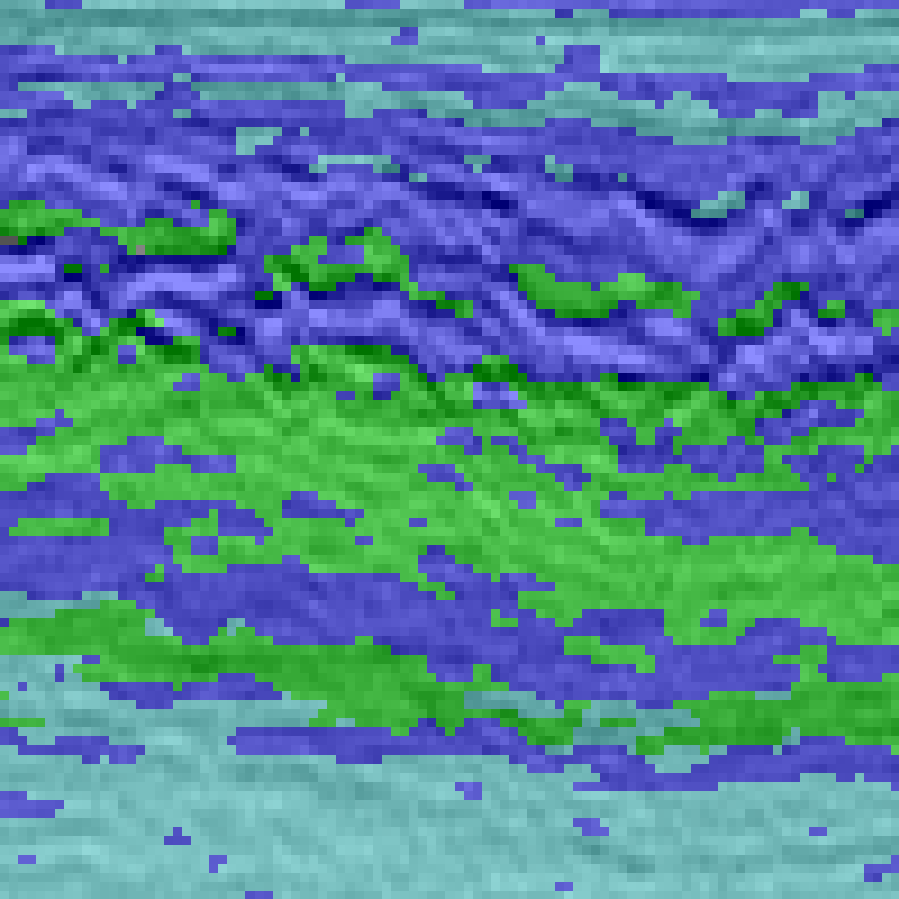}
&
\includegraphics[width=1.55cm]{figures/images/img_994.png} 
\includegraphics[width=1.55cm]{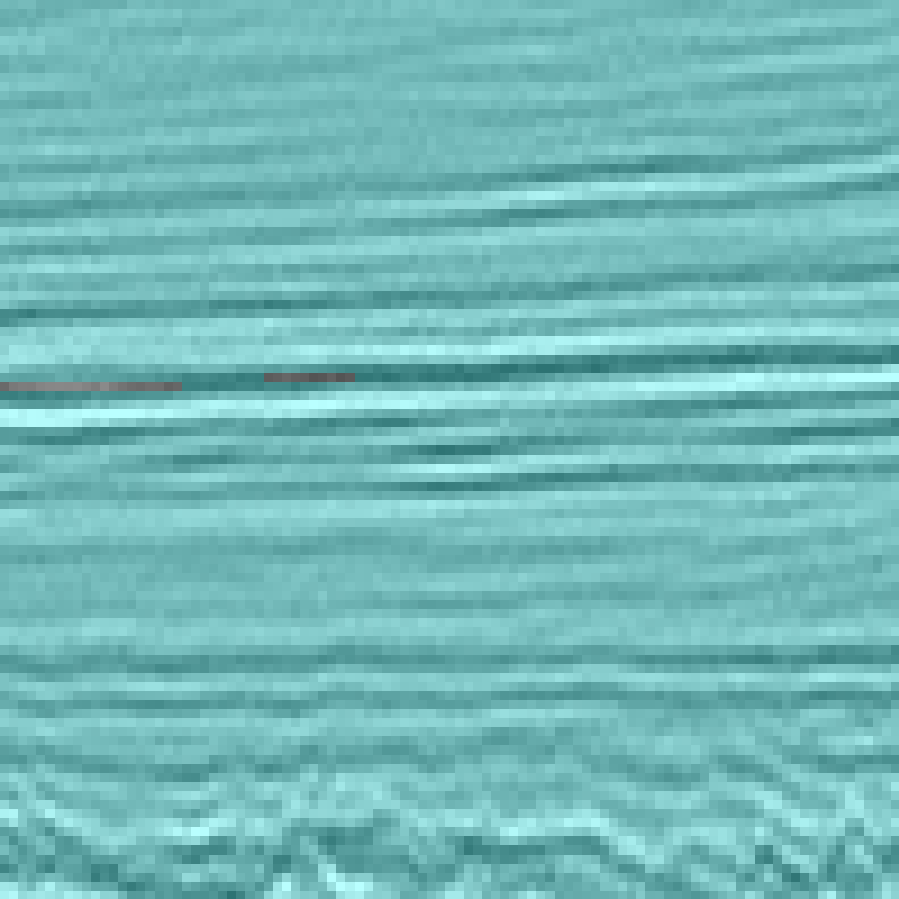}
&
\includegraphics[width=1.55cm]{figures/images/img_1436.png} 
\includegraphics[width=1.55cm]{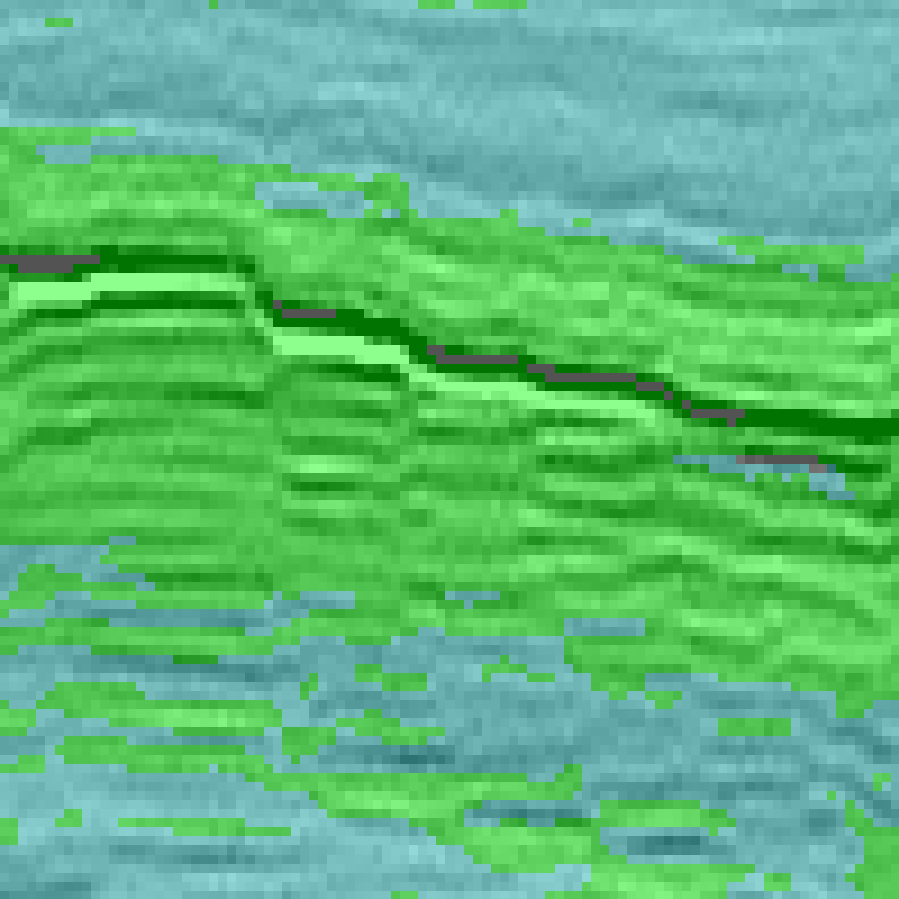}
&
\includegraphics[width=1.55cm]{figures/images/img_1897.png} 
\includegraphics[width=1.55cm]{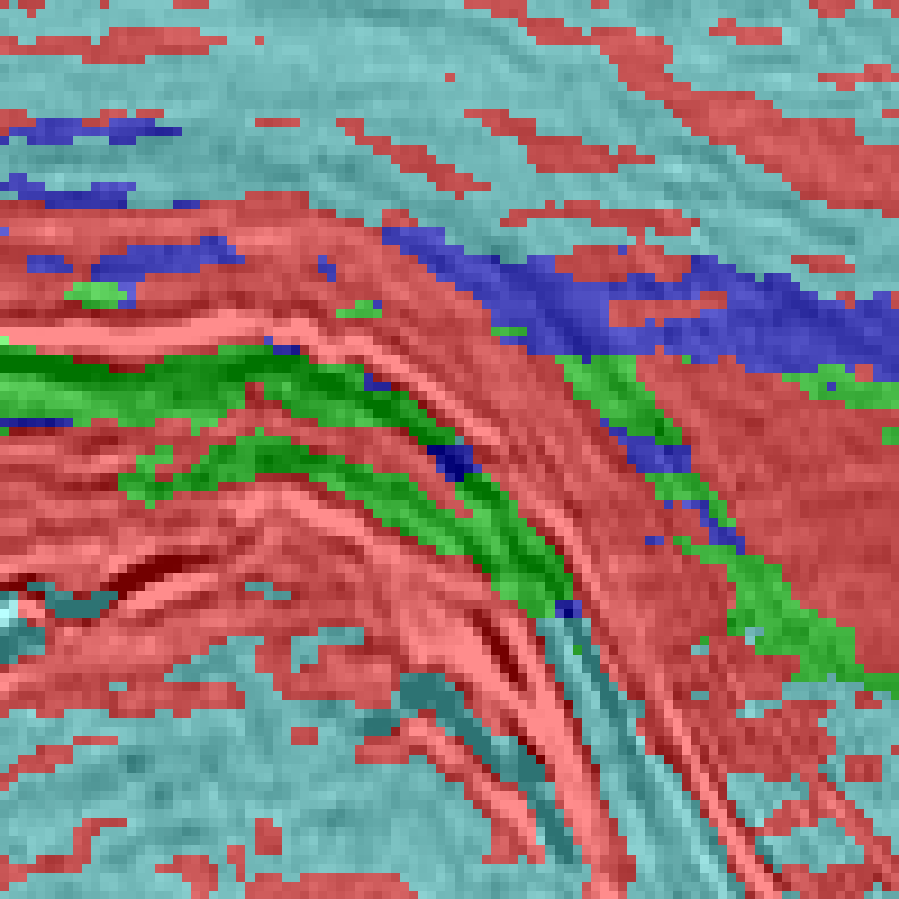}
\\
& & & 
\\
\multirow{3}{*}{\STAB{\rotatebox[origin=c]{90}{\footnotesize Proposed~~~~}}}
&
\includegraphics[width=1.55cm]{figures/images/img_1.png} 
\includegraphics[width=1.55cm]{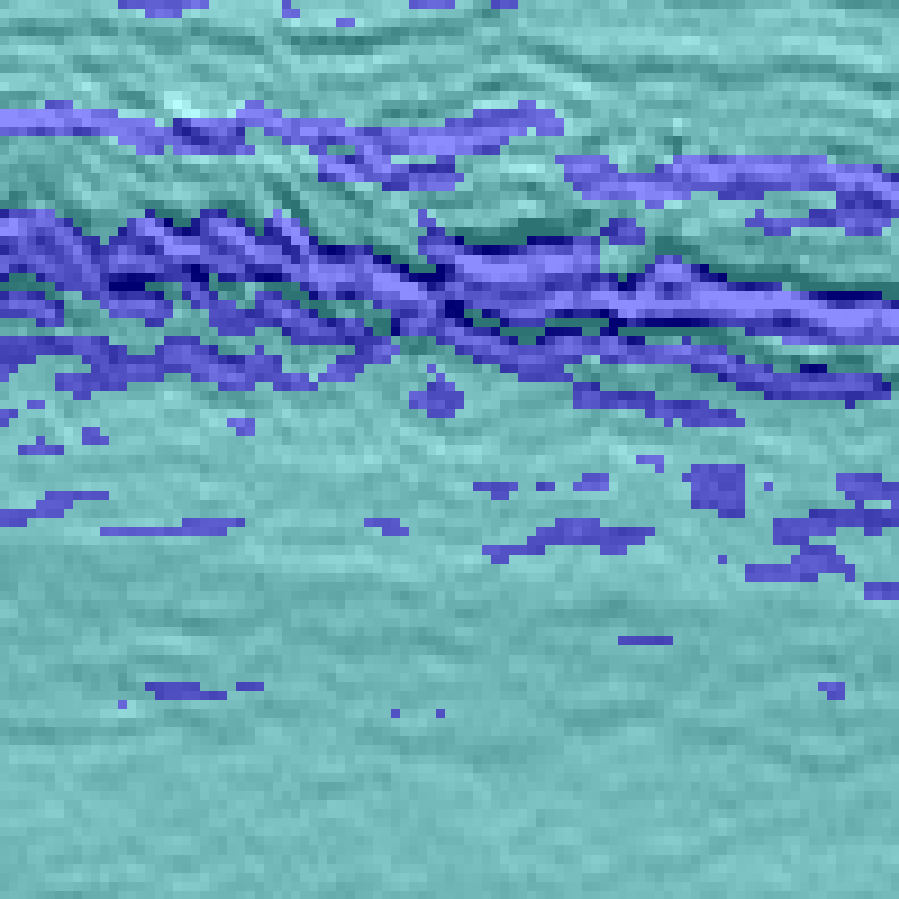}
&
\includegraphics[width=1.55cm]{figures/images/img_520.png} 
\includegraphics[width=1.55cm]{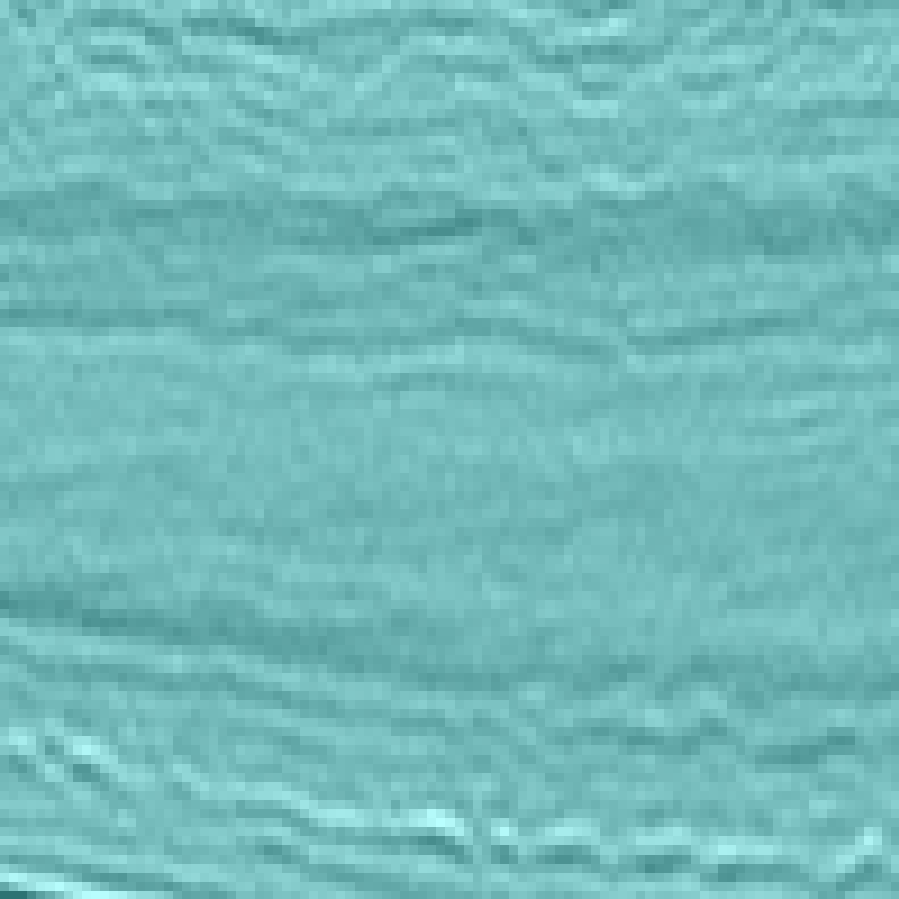}
&
\includegraphics[width=1.55cm]{figures/images/img_1026.png} 
\includegraphics[width=1.55cm]{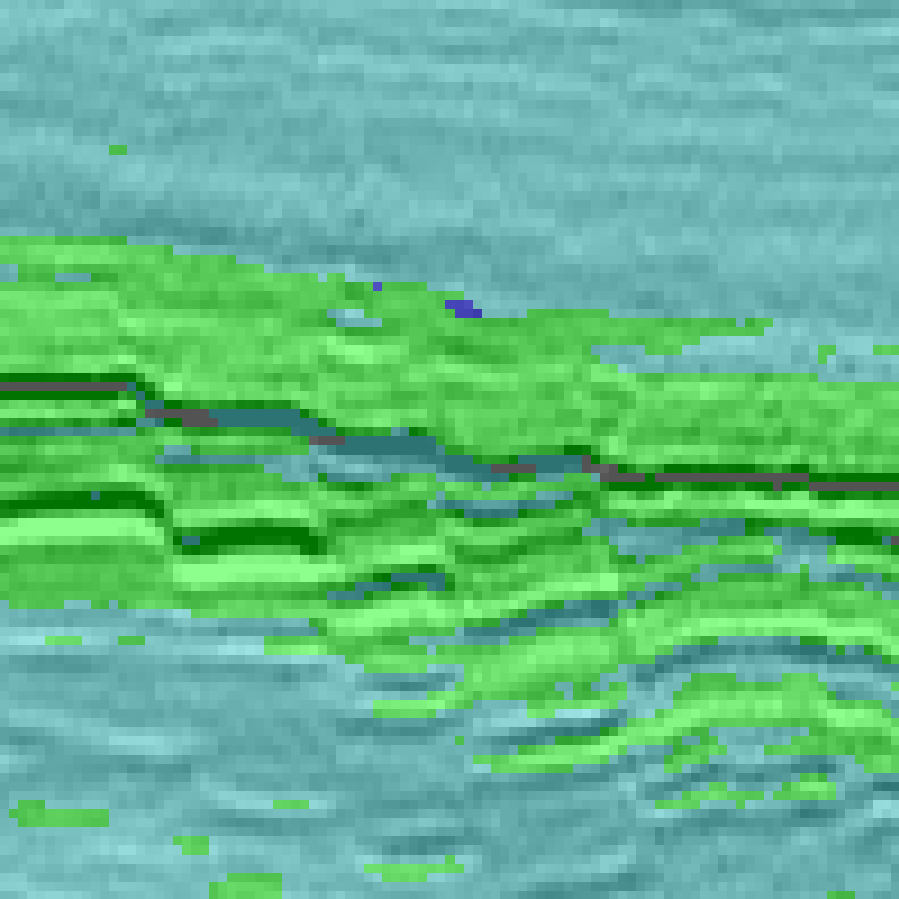}
&
\includegraphics[width=1.55cm]{figures/images/img_1539.png} 
\includegraphics[width=1.55cm]{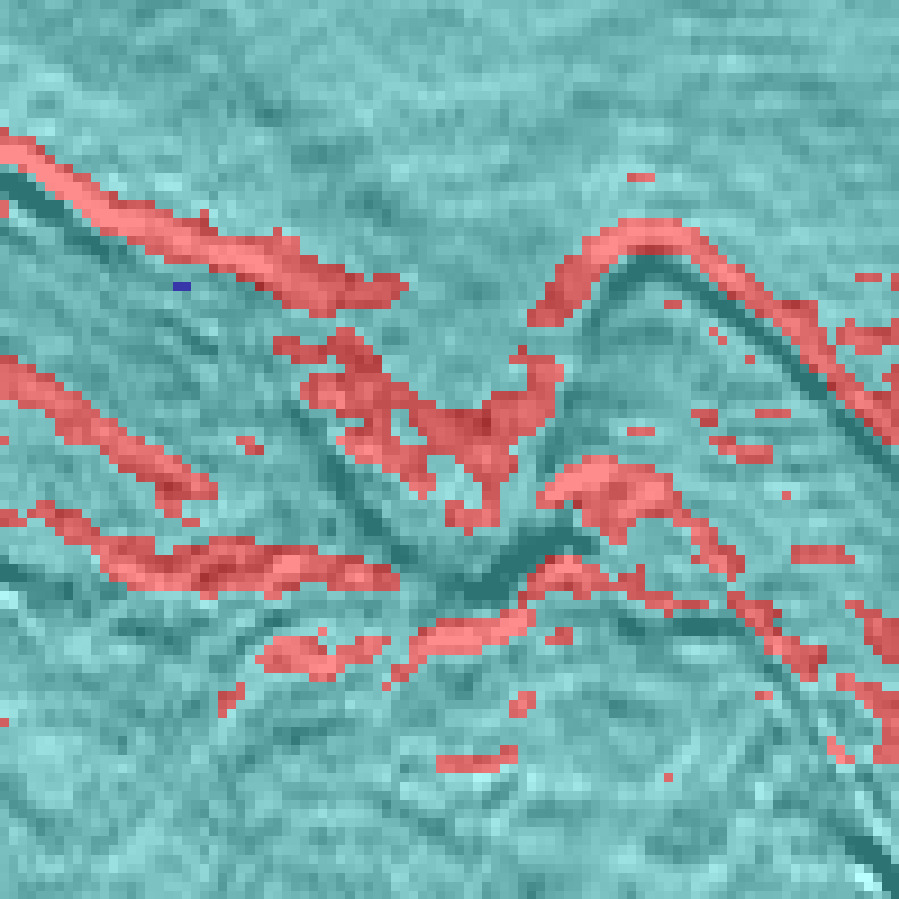}
\\
&
\includegraphics[width=1.55cm]{figures/images/img_104.png} 
\includegraphics[width=1.55cm]{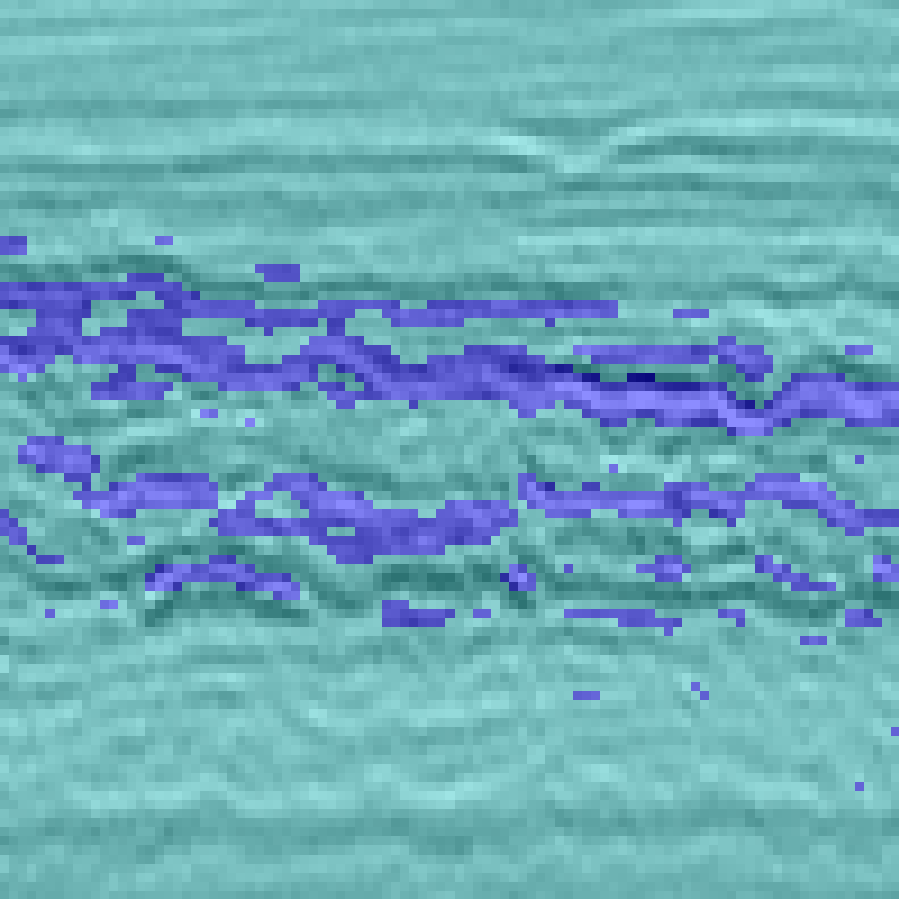}
&
\includegraphics[width=1.55cm]{figures/images/img_527.png} 
\includegraphics[width=1.55cm]{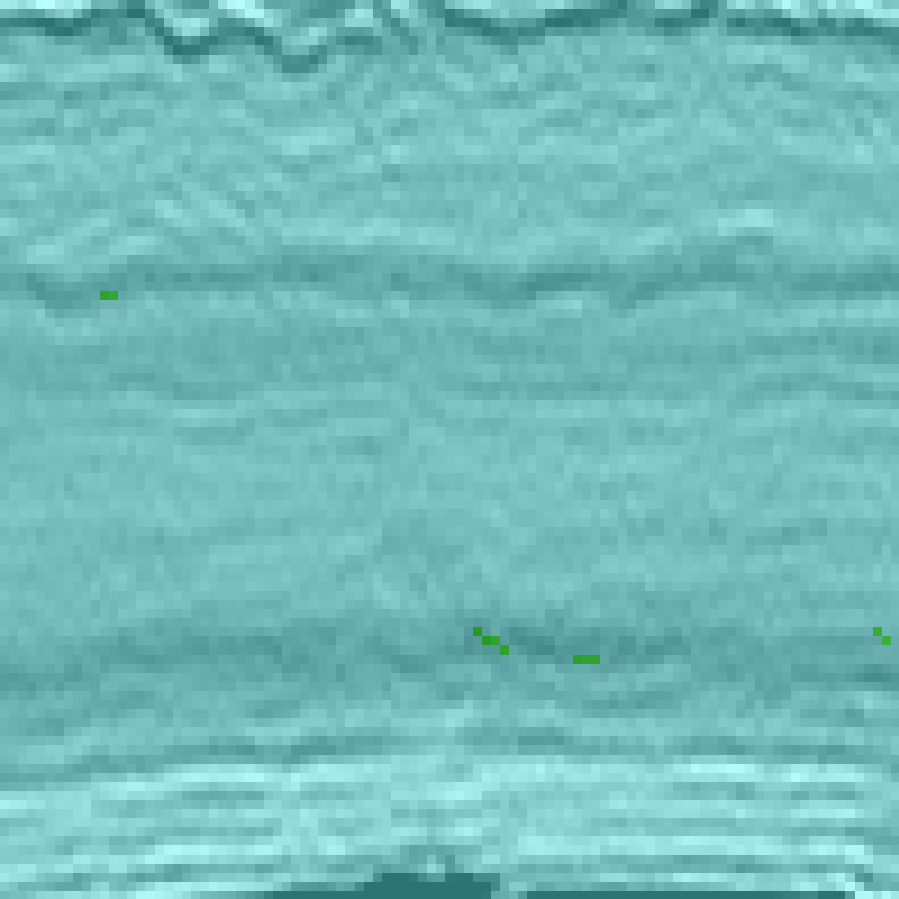}
&
\includegraphics[width=1.55cm]{figures/images/img_1129.png} 
\includegraphics[width=1.55cm]{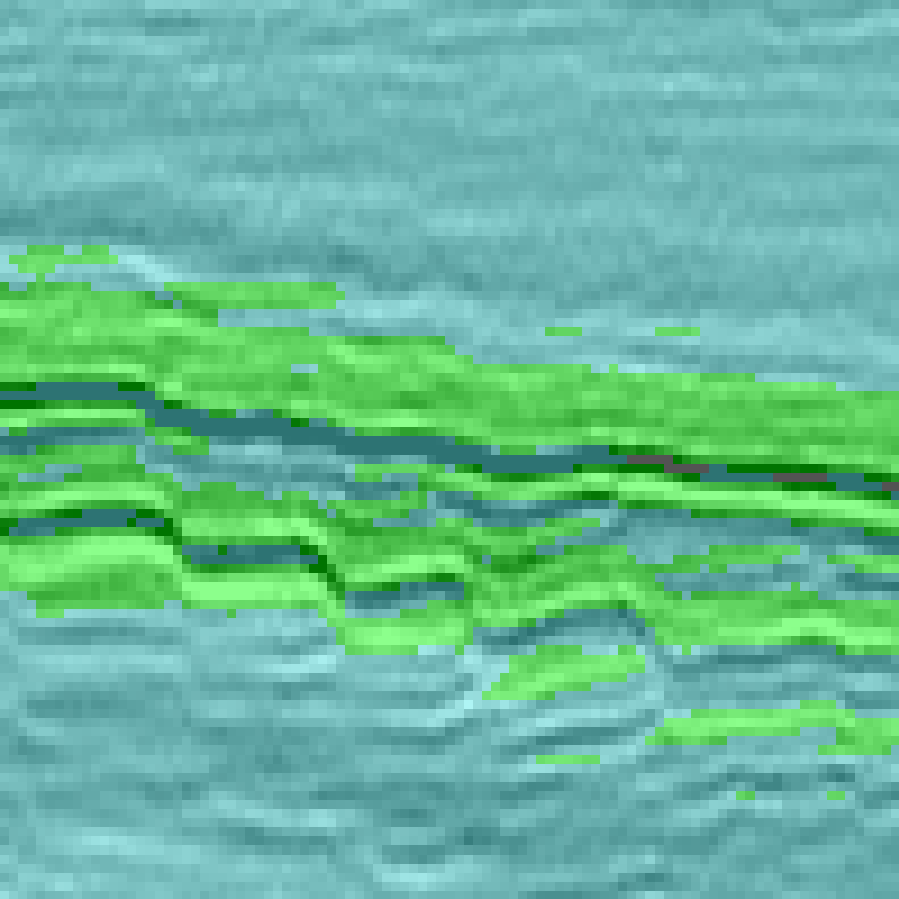}
&
\includegraphics[width=1.55cm]{figures/images/img_1638.png} 
\includegraphics[width=1.55cm]{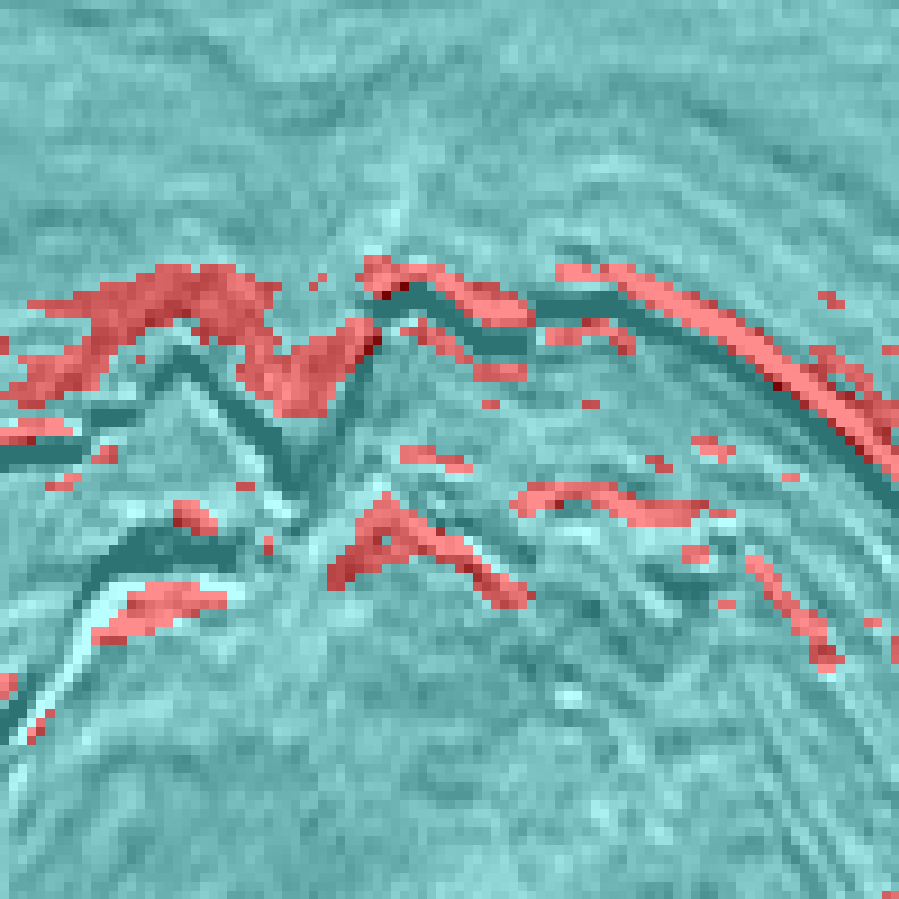}
\\
&
\includegraphics[width=1.55cm]{figures/images/img_360.png} 
\includegraphics[width=1.55cm]{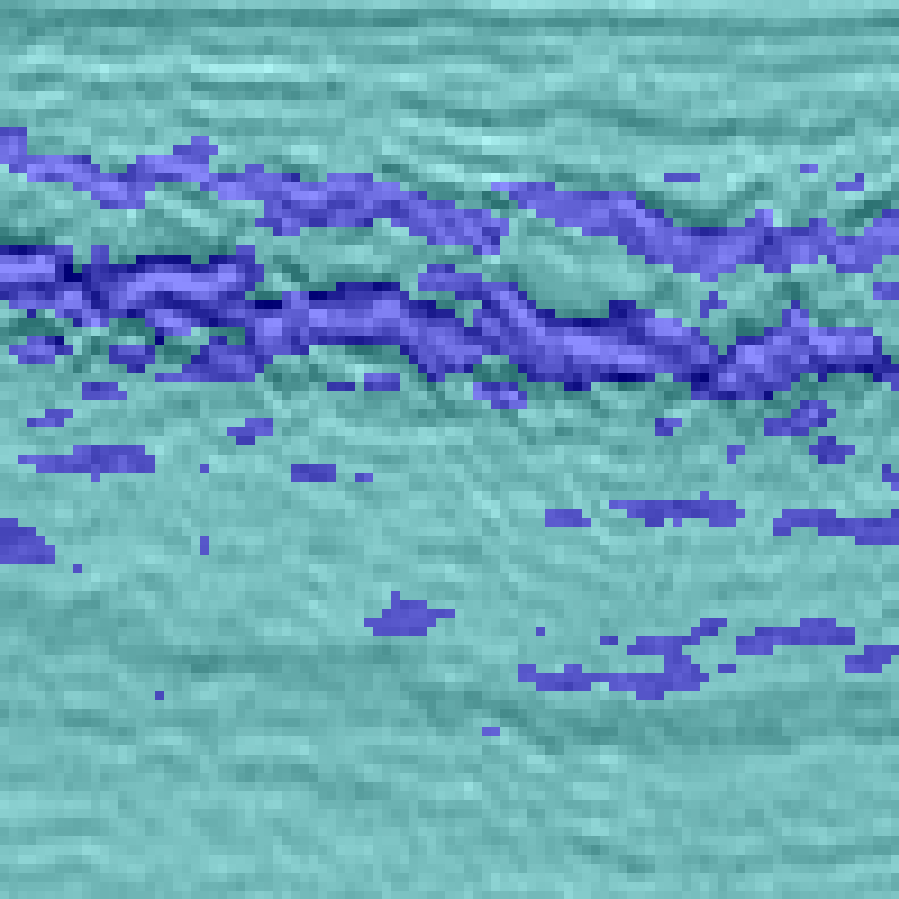}
&
\includegraphics[width=1.55cm]{figures/images/img_994.png} 
\includegraphics[width=1.55cm]{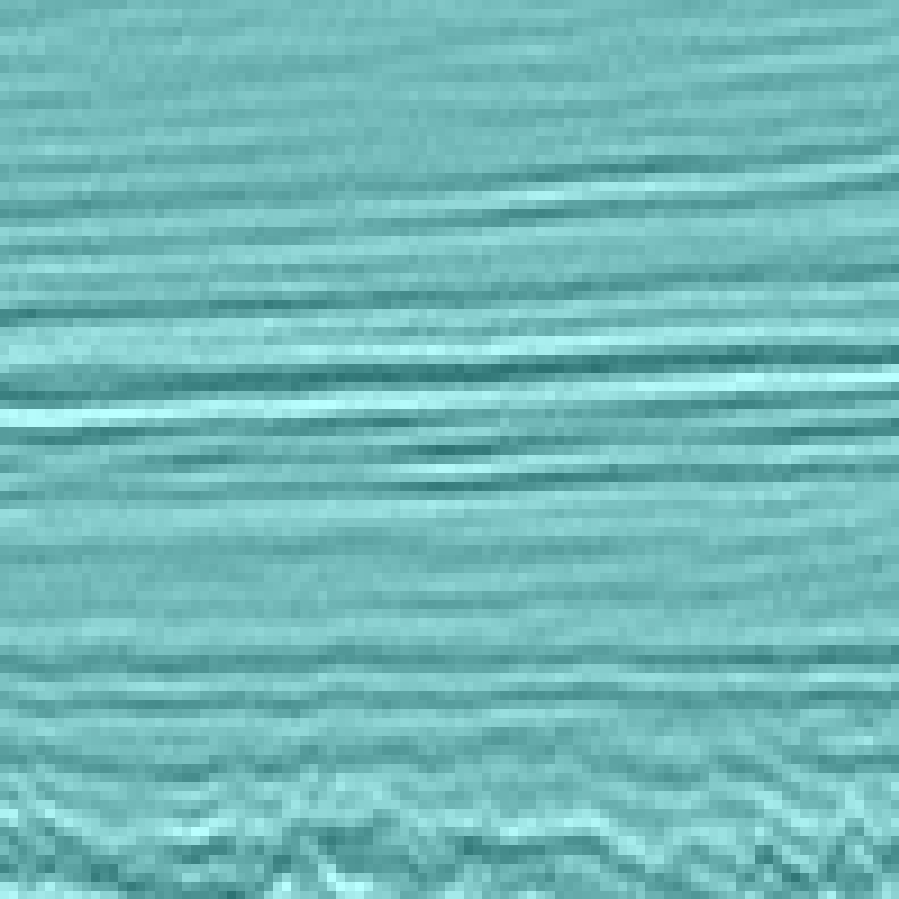}
&
\includegraphics[width=1.55cm]{figures/images/img_1436.png} 
\includegraphics[width=1.55cm]{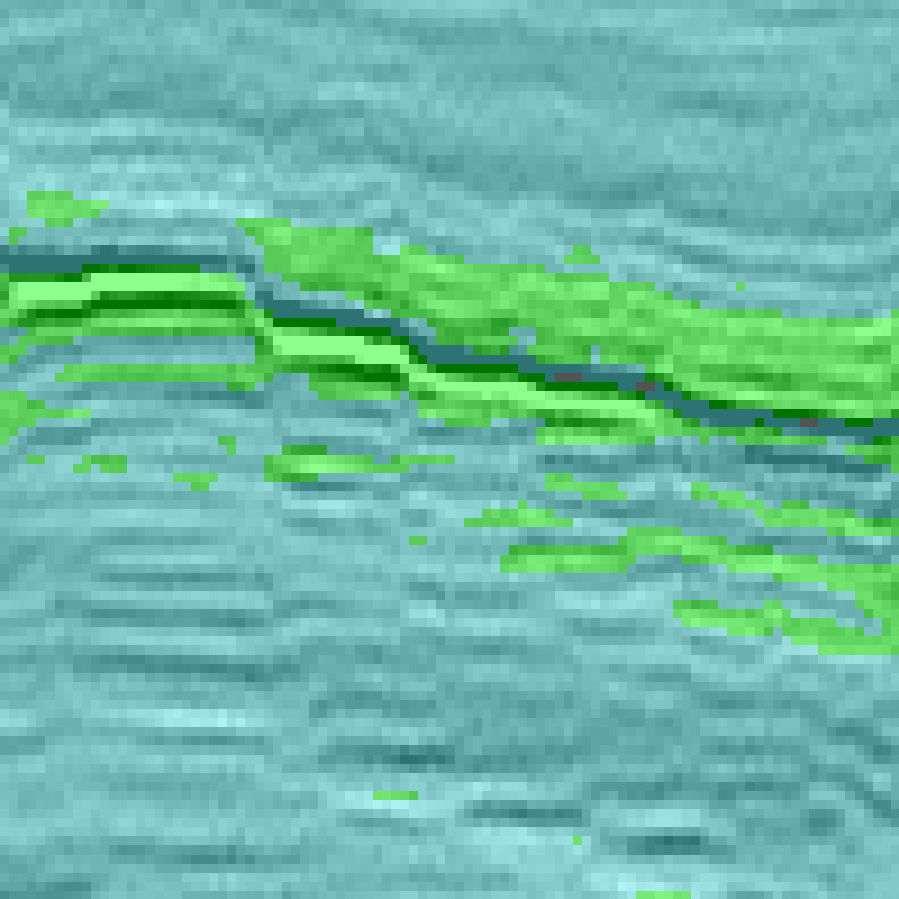}
&
\includegraphics[width=1.55cm]{figures/images/img_1897.png} 
\includegraphics[width=1.55cm]{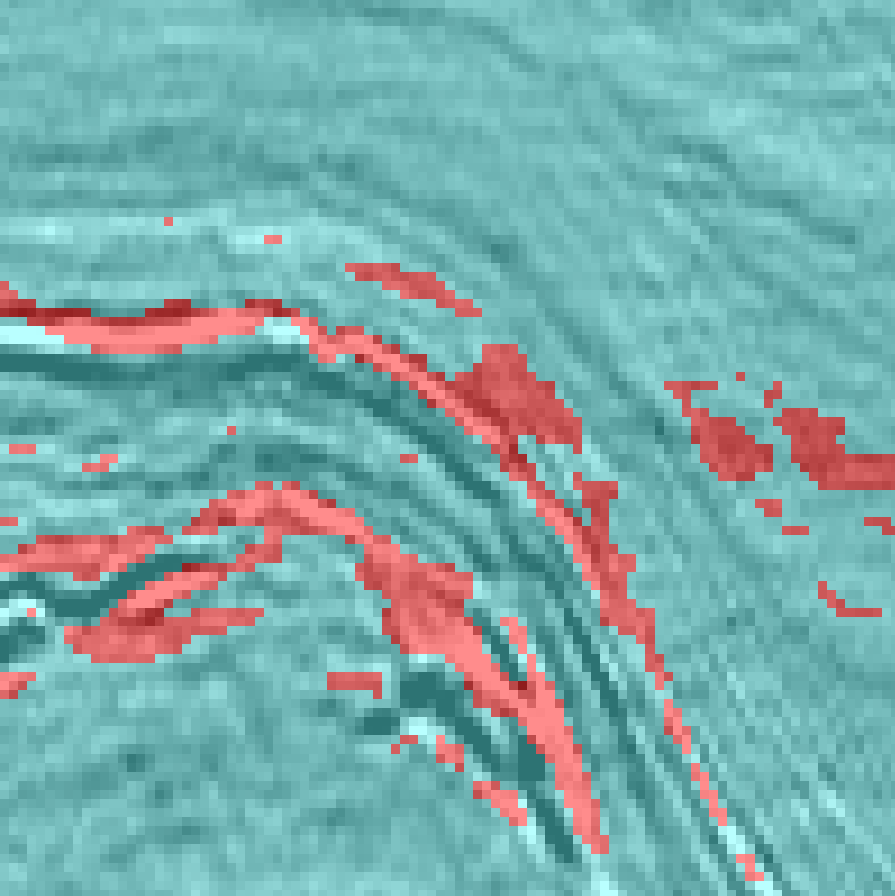}
\\
\end{tabular}
\end{center}

  \caption{A small sample of the final results of the weakly-supervised label mapping algorithm using NMF (top), NMF and orthogonality constraint on $\mathbf{H}$ (middle), and our proposed formulation (bottom). The first two columns show images containing \texttt{chaotic} structures, and the corresponding \texttt{chaotic} pixel-level labels generated by our method in blue. The middle two columns show images that contain \texttt{fault} structures, and \texttt{fault} pixel-level labels generated by our method in green. The last two columns show images that contain \texttt{salt dome} boundaries, and \texttt{salt dome} pixel-level labels generated by our method in red. Cyan corresponds to the \texttt{other} class.}
  \label{fig:results}
\end{figure*}

Finally, as with any other workflow, there are several areas where this approach can be improved. First, different classes of subsurface structures can often have different scales, whereas the method we have currently proposed uses a fixed size image for every class. It is worth investigating methods to alleviate this issue. Also, the final pixel-level labels are sensitive to the initial features, $\mathbf{W}^0$.  While we have showed that $k$-means can easily be used to initialize $\mathbf{W}^0$, it is worth investigating other more promising methods such as convolutional neural networks. In addition, if the data matrix  $\mathbf{X}$ has a wrong sparsity structure, applying the sparsity constraint in equation \ref{eq:sparsity} to form the feature matrix $\mathbf{W}$ might not lead to representative features of the different classes in $\mathbf{X}$. In that case, other techniques should be used to initialize $\mathbf{W}$. Finally, there are a few parameters such as the sparsity level $\rho_w$, the number of retrieved images per class $M$, and the regularization constants that need to be set by the interpreter based on her empirical assessment of the results. 

\section{Conclusion}

In conclusion, we have proposed a method for predicting the labels of various seismic structures to enable the use of fully-supervised machine learning techniques for seismic interpretation. Our method uses a few exemplar images that are assigned image-level labels by an interpreter. Thousands of visually similar images that contain the same structures are then automatically retrieved. Furthermore, our weakly-supervised label mapping algorithm learns the joint structures in all these retrieved images and maps their image-level labels into pixel-level labels that are more suitable for fully supervised machine learning applications. We applied our method to data extracted from the Netherlands Offshore F3 block. Our results show that this approach accurately labels the locations of subsurface structures such as faults, chaotic horizons, and salt dome boundaries using at most two exemplar images for each class.

This approach can also be used to predict labels for facies classification and similar problems within seismic interpretation. We believe this approach can significantly reduce the time and effort needed to obtain quality labeled data for training supervised machine learning models for seismic interpretation tasks. We consider that this work, and others in the future, will help open the way for more machine learning advances in seismic interpretation, and structural interpretation in particular.  

\section{Acknowledgments}
The authors would like to acknowledge the support of the Center for Energy and Geo Processing (CeGP) at Georgia Institute of Technology in Atlanta, USA, and King Fahd University of Petroleum and Minerals (KFUPM) in Dhahran, Saudi Arabia.

\append{Performance Evaluation Metrics}
\label{appendix:metrics}

We use several metrics to evaluate the performance of our method in the retrieval and clustering experiments mentioned in the results section. Here, we will explain these metrics in detail.

\subsubsection{Retrieval Evaluation Metrics}
To assess the retrieval performance of our method compared to other methods in the literature, we set each image $\mathbf{x}_i$ in the dataset as a query and retrieve the top images in terms of their similarity values. The performance of a similarity measure is quantified using information retrieval metrics. In order to present these metrics, let us first define the following sets: 
\begin{itemize}

    \item $\mathcal{R}_i^{(j)} = \left\{\mathbf{r}_i^{(1)},\mathbf{r}_i^{(2)},\dots,\mathbf{r}_i^{(j)} \right\}$ is the set of the first $j$ retrieved images for $\mathbf{x}_i$. Note that the elements of $\mathcal{R}_i^{(j)}$ are sorted according to their similarity to $\mathbf{x}_i$ such that $\mathsf{Similarity}\left(\mathbf{x}_i,\mathbf{r}_i^{(k)}\right)\geq \mathsf{Similarity}\left(\mathbf{x}_i,\mathbf{r}_i^{(k+1)}\right) $. 
    
    \item $\mathcal{C}_i$ is the set of all images that are of the same class as $\mathbf{x}_i$; excluding the image itself.
    \item $\mathcal{R}_i^{(j)} \cap \mathcal{C}_i$ is the intersection set of $\mathcal{R}_i^{(j)}$  and $\mathcal{C}_i$. It contains images that are of the same class as the query image $\mathbf{x}_i$ in the set of retrieved images $\mathcal{R}_i^{(j)}$.
    
    \end{itemize}

\noindent Next, we define information retrieval metrics that were used to assess the performance of the similarity measures. 

\begin{itemize}
    \item \textbf{Precision at $M$ (P@$M$)} is the average percentage of the correctly retrieved images when $M$ images are retrieved. Formally, 
    \begin{equation}
       \text{P@$M$} = \frac{1}{N_s} \sum_{i=1}^{N_s} \frac{\left\lvert \mathcal{R}_i^{(n)}\cap \mathcal{C}_i \right\rvert}{\left\lvert \mathcal{R}_i^{(n)} \right\rvert}, 
    \end{equation}
where $| \cdot |$ is the number of elements in the set.

\item \textbf{Retrieval Accuracy (RA)} is the P@$n$ when $n$ is equal to the number of elements that are of the same class the query images, i.e. $n=\lvert\mathcal{C}_i\rvert$. 
    \begin{equation}
      \text{RA} = \frac{1}{N_s} \sum_{i=1}^{N_s} \frac{\left\lvert \mathcal{R}_i^{(\lvert \mathcal{C}_i\rvert)}\cap \mathcal{C}_i \right\rvert}{\left\lvert \mathcal{R}_i^{(\lvert \mathcal{C}_i\rvert)} \right\rvert}. 
    \end{equation}

  \item \textbf{Average Precision (AP)} for query image $\mathbf{x}_i$ is a measure of precision that takes into account the order of which the correct images are retrieved. It is defined as: 
        \begin{equation}
        \text{AP}_i = \frac{1}{\left\lvert \mathcal{C}_i \right\rvert} \sum_{j=1}^{N_s-1} \frac{\left\lvert \mathcal{R}_i^{(j)}\cap \mathcal{C}_i \right\rvert}{\left\lvert \mathcal{R}_i^{(j)} \right\rvert}\times  {1}_{\{r_i^{(j)}\in~ \mathcal{C}_i\}},
        \end{equation}
    
where ${1}_{\{r_i^{(j)}\in~ \mathcal{C}_i\}}$ is the indicator function and it is equal to $1$ if and only if $r_i^{(j)}\in \mathcal{C}_i$, and $0$ otherwise. \textbf{Mean Average Precision (MAP)} is the mean value of AP for all images in the dataset. 

\item \textbf{Receiver Operating Characteristics (ROC)} is a plot of the True Positive Rate (TPR) versus False Positive Rate (FPR) for different similarity thresholds. TPR is the percentage of pairs of images that are correctly identified as similar by the similarity measure. FPR is the percentage of pairs of images that are not similar but were identified as similar by the similarity measure. The area under the ROC curve, denoted as \textbf{AUC}, is used as a measure of the discriminative power of a similarity measure. The ideal ROC curve would have perfect TPR (TPR=1) for all values of FPR, and in this case, the area under the curve would be maximum AUC = 1.

\end{itemize}

\subsubsection{Clustering Evaluation Metric}

We use the \textbf{rand index} to evaluate our performance in the clustering experiments. The rand index is defined as follows. For each pair of images, $\mathbf{x}_i$ and $\mathbf{x}_j$, in the dataset, we compare the results obtained by $k$-means clustering with the ground truth which are the image labels. Then we count the number of correctly clustered pairs. A pair is said to be correctly clustered if: 
\begin{itemize}
\item $\mathbf{x}_i$ and $\mathbf{x}_j$ are of the same class and are in the same cluster in the similarity-based clustering. 

\item $\mathbf{x}_i$ and $\mathbf{x}_j$ are of different classes and are in different clusters in the similarity-based clustering. 
\end{itemize}

\noindent If $p_\text{correct}$ is the total number of correctly clustered pairs and $p_\text{total} = \binom{N_s}{2}$ is the total number of possible pairs in the dataset, The Rand Index is defined as the ratio of the two numbers, 

\begin{equation}
    \text{\textbf{Rand Index}} = \frac{p_\text{correct}}{p_\text{all}} = \frac{2p_\text{correct}}{N_s (N_s-1)}.
\end{equation}

\noindent All of the metrics defined above are in the range $[0,1]$ with $1$ being a perfect score. 

\append{Derivation of the Multiplicative Update Rules}
\label{appendix:mur}

We would like to derive the multiplicative update rules shown in equations \ref{WMUR} and \ref{HMUR}. These multiplicative update rules solve the optimization problems introduced in equations \ref{eq:WObjective} and \ref{eq:HObjective}. We adopt an approach similar to that proposed by \cite{Lee2001}. Namely, we will derive the gradient descent updates to solve the problem for $\mathbf{W}$ and $\mathbf{H}$ separately. Then, we solve the problem in equation \ref{SONMF} by alternately updating $\mathbf{W}$ and $\mathbf{H}$ successively until they converge. We will derive the multiplicative update rules using the objective functions in equations \ref{eq:WObjective} and \ref{eq:HObjective}. For the sake of the simplicity of the derivation, we drop all the constraints in equations \ref{eq:WObjective} and \ref{eq:HObjective}, and later show that the derived multiplicative update rules for the non-constrained problem also solve the constrained optimization problem under the conditions that we have. Additionally, we have shown in Figure \ref{fig:optimization} that solving these two problems iteratively also solves the problem in equation \ref{SONMF}. Therefore, for matrix  $\mathbf{W}$ we have 
\begin{equation}\label{eq:W-no-cons}
\underset{\mathbf{W}}{\arg\min}  || \mathbf{X} - \mathbf{W}\mathbf{H} ||_F^2 + \lambda_1 ||\mathbf{W}||_F^2,
\end{equation}
\noindent
and for $\mathbf{H}$, we have
\begin{equation}\label{eq:H-no-cons}
\underset{\mathbf{H}}{\arg\min}  || \mathbf{X} - \mathbf{W}\mathbf{H} ||_F^2  +  \gamma||\mathbf{H}\mathbf{H}^T - \mathbf{I}||_F^2 + \lambda_2||\mathbf{H}||_F^2.  
\end{equation}

We derive the multiplicative update rules for $\mathbf{W}$ and $\mathbf{H}$ respectively in the following two subsections. 

\subsection{Multiplicative Update Rule for $\mathbf{W}$}

If we denote the objective function defined in \ref{eq:W-no-cons} as $\mathcal{F}_\mathbf{W}$, we can rewrite $\mathcal{F}_\mathbf{W}$ as 

\begin{equation}
\mathcal{F}_\mathbf{W} = \mathrm{Tr}\big( (\mathbf{X} - \mathbf{W}\mathbf{H})^T(\mathbf{X} - \mathbf{W}\mathbf{H}) \big) + \lambda_1 \mathrm{Tr}(\mathbf{W}^T\mathbf{W}),
\end{equation}
\noindent where $\mathrm{Tr}( \cdot )$ denotes the trace of a matrix. Simplifying the expression further, and employing the property that $\mathrm{Tr}(\mathbf{A} + \mathbf{B}) = \mathrm{Tr}(\mathbf{A})  + \mathrm{Tr}(\mathbf{B})$, we obtain
\begin{equation}
\begin{aligned}
\mathcal{F}_\mathbf{W} &= \mathrm{Tr}(\mathbf{X}^T\mathbf{X}) - \mathrm{Tr}(\mathbf{X}^T\mathbf{W}\mathbf{H}) - \mathrm{Tr}(\mathbf{H}^T\mathbf{W}^T\mathbf{X})  \\
&+\mathrm{Tr}(\mathbf{H}^T\mathbf{W}^T \mathbf{W}\mathbf{H}) +\lambda_1 \mathrm{Tr}(\mathbf{W}^T\mathbf{W}).
\end{aligned}
\end{equation}
\noindent Taking the partial derivative of $\mathcal{F}_\mathbf{W}$ with respect to $\mathbf{W}$ we get
\begin{equation}
\begin{aligned}
\frac{\partial \mathcal{F}_\mathbf{W}}{\partial \mathbf{W}} &= -2(\mathbf{X}\mathbf{H}^T) + 2(\mathbf{W}\mathbf{H}\mathbf{H}^T) + 2\lambda_1 \mathbf{W} \\
&\propto ~ - \mathbf{X}\mathbf{H}^T +  \mathbf{W}\mathbf{H}\mathbf{H}^T + \lambda_1 \mathbf{W}  
\end{aligned}
\end{equation}
\noindent The gradient descent update for $\mathbf{W}$ will then be a step in the direction of the negative gradient. In other words, 
\begin{equation}\label{eq:A5}
    \mathbf{W}^{t+1} =  \mathbf{W}^t + \eta \big(\mathbf{X}{\mathbf{H}^t}^T - \mathbf{W}^t\mathbf{H}^t{\mathbf{H}^t}^T - \lambda_1 \mathbf{W}^t \big),  
\end{equation}

\noindent where $\eta$ is the step size. Note that this is an additive update rule. The negative signs indicate that even if the values in $\mathbf{X}$, $\mathbf{W}^0$ and $\mathbf{H}^0$ are non-negative, we are not guaranteed to arrive at a non-negative final solution. However, by selecting our step size as 
\begin{equation}
\eta = \frac{\mathbf{W}^t}{\mathbf{W}^t\mathbf{H}^t{\mathbf{H}^t}^T + \lambda_1 \mathbf{W}^t},
\end{equation}
\noindent and substituting in the gradient descent update in equation \ref{eq:A5}, the additive rule becomes a multiplicative update rule: 
\begin{equation}\label{eq:A7}
\mathbf{W}^{t+1} = \mathbf{W}^t \odot \frac{\big(\mathbf{X}{ \mathbf{H}^t}^T + \epsilon )_{ij}}{\big(\mathbf{W}^t\mathbf{H}^t{\mathbf{H}^t}^T + \lambda_1 \mathbf{W}^t + \epsilon \big)_{ij}}.
\end{equation}
\noindent We add a small positive real number $\epsilon$ to avoid division by zero. This result is identical to the result in equation \ref{WMUR}. 

\subsection{Multiplicative Update Rule for $\mathbf{H}$}
\raggedbottom
Similarly for $\mathbf{H}$, we write the objective function in equation \ref{eq:H-no-cons} as

\begin{equation}
\begin{aligned}
\mathcal{F}_\mathbf{H} &= \mathrm{Tr}\big( (\mathbf{X} - \mathbf{W}\mathbf{H})^T(\mathbf{X} - \mathbf{W}\mathbf{H}) \big) 
+\lambda_2 \mathrm{Tr}(\mathbf{H}^T\mathbf{H}) \\ &+\gamma \mathrm{Tr}\big((\mathbf{H}\mathbf{H}^T - \mathbf{I})^T(\mathbf{H}\mathbf{H}^T - \mathbf{I})\big).
\end{aligned}
\end{equation}
Taking the partial derivative of $\mathcal{F}_\mathbf{H}$ with respect to $\mathbf{H}$, and simplifying the expression further, 
\begin{equation}
\begin{aligned}
\frac{\partial \mathcal{F}_\mathbf{H}}{\partial \mathbf{H}} &= - 2\mathbf{W}^T\mathbf{X} + 2(\mathbf{W}^T\mathbf{W}\mathbf{H}) + 2 \lambda_2 \mathbf{H}\\
&+ 4 \gamma \mathbf{H}^T\mathbf{H} \mathbf{H}^T - 2 \gamma \mathbf{H}\\
&\propto  (\mathbf{W}^T\mathbf{W}\mathbf{H}) + \lambda_2 \mathbf{H}
+ \gamma \mathbf{H}^T\mathbf{H} \mathbf{H}^T \\&- \big( \mathbf{W}^T\mathbf{X} + \gamma \mathbf{H}\big).
\end{aligned}
\end{equation}
\noindent The gradient descent update step then becomes
\begin{equation}\label{eq:A9}
\begin{aligned}
    \mathbf{H}^{t+1} &= \mathbf{H}^t + \eta \Big(  {\mathbf{W}^{t+1}}^T\mathbf{X} + \gamma \mathbf{H}^t \\ 
    &- ({\mathbf{W}^{t+1}}^T{\mathbf{W}^{t+1}}\mathbf{H}^t + \lambda_2 \mathbf{H}^t
 + \gamma {\mathbf{H}^t}^T\mathbf{H}^t {\mathbf{H}^t}^T)
    \Big).
\end{aligned}
\end{equation}
\noindent If we select the step size to be
\begin{equation}
\eta = \frac{\mathbf{H}^t}{\mathbf{{W}^{t+1}}^T\mathbf{W}^{t+1}\mathbf{H}^t + \lambda_2 \mathbf{H}^t
 + \gamma{ \mathbf{H}^t}^T\mathbf{H}^t{ \mathbf{H}^t}^T}, 
\end{equation}
\noindent and substitute this value in equation \ref{eq:A9} and simplify, we arrive at the multiplicative update rule for $\mathbf{H}$
\begin{equation}\label{eq:A11}
\mathbf{H}^{t+1} = \frac{\mathbf{H}^t \odot \big( {\mathbf{W}^{t+1}}^T\mathbf{X}+\gamma \mathbf{H}^t + \epsilon  \big)_{ij}}{\big({\mathbf{W}^{t+1}}^T\mathbf{W}^{t+1}\mathbf{H}^t + \lambda_2 \mathbf{H}^t + \gamma {\mathbf{H}^t}^T\mathbf{H}^t{\mathbf{H}^t}^T + \epsilon \big)_{ij}}
\end{equation}
\noindent This is identical to the update rule shown in equation \ref{HMUR}. 

\subsection{Constrained Optimization}
The update rules in equations \ref{eq:A7} and \ref{eq:A11} solve the non-constrained problems in equation \ref{eq:W-no-cons} and \ref{eq:H-no-cons}. However, our original problem in equation \ref{SONMF} is a constrained one. Since we initialize the matrices $\mathbf{W}$ and $\mathbf{H}$ with non-negative values, it is trivial to see that the multiplicative update rules in equations \ref{eq:A7} and \ref{eq:A11} will always give non-negative results, thus satisfying the non-negativity constraint. Furthermore, since the sparsity constraint on the features $\mathbf{w}_i$ is applied to the initial features, $\mathbf{W}^0$, any further application of the update rule in equation \ref{eq:A7} will not modify the zero elements in the matrix $\mathbf{W}$, and hence, the initial feature sparsity is preserved. Therefore, although we solved for the non-constrained problem in equations \ref{eq:W-no-cons} and \ref{eq:H-no-cons}, our solution is still valid for the constrained problem in equation \ref{SONMF}. It is important to note that the final solution is dependant on the initial features $\mathbf{W}^0$. If $\mathbf{W}^0$ had a wrong sparsity structure, then the final results would not be accurate. Therefore, it is essential to initialize $\mathbf{W}^0$ properly to obtain accurate results. 

\bibliographystyle{seg} 
\bibliography{main.bib}

\end{document}